\documentclass[preprint2]{aastex}
\usepackage{graphicx}
\usepackage{subfigure}

\newcommand{\be}{\begin{equation}}
\newcommand{\ee}{\end{equation}}
\newcommand{\bea}{\begin{eqnarray}}
\newcommand{\eea}{\end{eqnarray}}
\newcommand{\gtabouteq}{\,\hbox{\raise 0.5 ex \hbox{$>$}\kern-.77em 
                    \lower 0.5 ex \hbox{$\sim$}$\,$}}       
\newcommand{\ltabouteq}{\,\hbox{\raise 0.5 ex \hbox{$<$}\kern-.77em 
    \lower 0.5 ex \hbox{$\sim$}$\,$}}
\def\rasec {\hbox{$\,$\raise 0.6 ex \hbox{\rm s}\kern-.35em
                  \lower 0.0 ex \hbox{.}$\,$}}        
\def\decsec{\hbox{$\,$\raise 0.5 ex \hbox{$\prime\prime$}\kern-.45em
                  \lower 0.0 ex \hbox{.}$\,$}}         
\def\decmin{\hbox{$\,$\raise 0.5 ex \hbox{$\prime$}\kern-.45em
    \lower 0.0 ex \hbox{}$\,$}}

\DeclareRobustCommand{\orderof}{\ensuremath{\mathcal{O}}}

\shorttitle{Circular Polarization in the CHANG-ES Sample}
\shortauthors{Irwin et al.}

\begin{document}


\title{CHANG-ES XI: \\
Circular Polarization in the Cores of Nearby Galaxies}


\author{
  Judith A. Irwin\altaffilmark{1},
  Richard N. Henriksen\altaffilmark{1},
Marek We{\.z}gowiec\altaffilmark{2},
  Ancor Damas-Segovia\altaffilmark{3},
  Q. Daniel Wang\altaffilmark{4},
  Marita Krause\altaffilmark{3},
  George Heald\altaffilmark{5},
Ralf-J{\"u}rgen Dettmar\altaffilmark{6},
Jiang-Tao Li\altaffilmark{7},
Theresa Wiegert\altaffilmark{1},
Yelena Stein\altaffilmark{6},
Timothy T. Braun\altaffilmark{8},
Jisung Im\altaffilmark{1},
Philip Schmidt\altaffilmark{3},
Scott Macdonald\altaffilmark{1},
Arpad Miskolczi\altaffilmark{6},
Alison Merritt\altaffilmark{1},
S. C. Mora-Partiarroyo\altaffilmark{3},
D. J. Saikia\altaffilmark{9},
Carlos Sotomayor\altaffilmark{6},
Yang Yang\altaffilmark{10}
}
\altaffiltext{1}{Dept. of Physics, Engineering Physics, \& Astronomy, Queen's University,
  Kingston, Ontario, Canada, K7L 3N6 {\tt irwinja@queensu.ca, henriksn@astro.queensu.ca, twiegert@queensu.ca, j4im@uwaterloo.ca, macdonscott@gmail.com, 11ajm24@queensu.ca}.}
\altaffiltext{2}{Obserwatorium Astronomiczne Uniwersytetu Jagiello\'nskiego, ul. Orla 171, 30-244 Krak\'ow, Poland
  {\tt markmet@oa.uj.edu.pl}.
}
\altaffiltext{3}{Max-Planck-Institut f{\"u}r Radioastronomie,  Auf dem H{\"u}gel 69,
53121, Bonn, Germany,
{\tt adamas@mpifr-bonn.mpg.de, mkrause@mpifr-bonn.mpg.de, pschmidt@mpifr-bonn.mpg.de,
 silvia.carolina.mora@gmail.com}.} 
\altaffiltext{4}{Dept. of Astronomy, University of Massachusetts, 710 North
Pleasant St., Amherst, MA, 01003, USA, 
{\tt wqd@astro.umass.edu}.}
\altaffiltext{5}{CSIRO Astronomy and Space Science, 26 Dick Perry Avenue, Kensington WA 6151, Australia {\tt George.heald@csiro.au}.}
\altaffiltext{6}{Astronomisches Institut, Ruhr-Universit{\"a}t Bochum, 44780 Bochum,
 Germany,
{\tt dettmar@astro.rub.de, stein@astro.rub.de, miskolczi@astro.rub.de, carlos.sotomayor@gmail.com}.}
\altaffiltext{7}{Department of Astronomy, University of Michigan, 311 West Hall, 1085 S. University Ave.,
Ann Arbor, MI, 48109 {\tt jiangtal@umich.edu}.}
\altaffiltext{8}{Dept. of Physics and Astronomy, University of New Mexico, 
1919 Lomas Boulevard, NE, Albuquerque, NM, 87131, USA {\tt ttbraun@unm.edu}.}
\altaffiltext{9}{National Centre for Radio Astrophysics, 
TIFR, Pune University Campus, Post Bag 3, Pune, 411 007, India,
{\tt djsaikia21jan@gmail.com; djs@ncra.tifr.res.in}.}
\altaffiltext{10}{Nanjing University, Nanjing, Jiangsu, 210093, China {\tt yangyang.astro@gmail.com; stfyou@163.com}.}




\begin{abstract}
We detect 5 galaxies in the CHANG-ES (Continuum Halos in Nearby Galaxies -- an EVLA Survey) sample that show circular polarization (CP) at L-band in our high resolution data sets. Two of the galaxies (NGC~4388 and NGC~4845) show strong Stokes $V/I\,\equiv\,m_C\,\sim\,2$\%,  two (NGC~660 and NGC~3628) have values of $m_C\sim \,0.3$\%, and NGC~3079 is a marginal  detection at $m_C\sim \,0.2$\%.  The two strongest $m_C$ galaxies also have the most luminous X-ray cores and the strongest internal absorption in X-rays.  
We have expanded on our previous Faraday conversion interpretation and analysis and provide analytical expressions for the expected $V$ signal for a general case in which the cosmic ray electron energy spectral index can take on any value.  We provide examples as to how such expressions could be used to estimate magnetic field strengths and the lower energy cutoff for CR electrons.  
  Four out of our detections are {\it resolved}, showing unique structures, including a {\it jet} in NGC~4388 and a CP `conversion disk' in NGC~4845. The conversion disk is inclined to the galactic disk but is perpendicular to a possible outflow direction.  Such CP structures have never before been seen in any galaxy to our knowledge. None of the galaxy cores show linear polarization at L-band.  Thus CP may provide a unique probe of physical conditions deep into radio AGNs. 
\end{abstract}





\section{Introduction}
\label{sec:introduction}

Circular polarization (hereafter {CP}) has been an elusive quantity to measure in extragalactic sources.  When present, it is a weak signal with a typical value of $m_C\sim$ 0.2\% or less in Stokes $|V/I|$, where $m_C\,\equiv\,{V/I}$ \citep[][]{ray00}\footnote{We distinguish between `CP' which refers to any detection of Stokes $V$ and the ratio, $m_C\,\equiv\,V/I$.}. A strong signal could be considered $>0.3\%$ \citep{hom06}.
In addition, observing CP is technically challenging (Sect.~\ref{sec:special}) and CP, when measured, can be highly variable (see below).


When observed,
the most successful mechanism for explaining CP and most closely matching the observations thus far, is {\it Faraday conversion} \citep[e.g.][]{jon88,mac02,bec02,bec03,ens03,osu13,irw15}.
Initially linearly polarized emission, when travelling through a birefringent medium, can be converted to CP, provided the angle between the linearly polarized signal and the magnetic field varies along the line of sight. This condition can be achieved by normal Faraday rotation in a regular $B$ field, or by a changing $B$ field such as might be seen in a jet with a helical $B$ component \citep{ens03,irw15}. 

CP is often observed in or near the optically thick regime in total intensity \citep{ray00}. 
 This may be because Stokes $V$ is predicted to have a steep negative spectral index in most cases (see Sect.~\ref{sec:predictions}).  Thus low frequency observations are more likely to result in a CP detection and this is where sources are more likely to be optically thick in total intensity.
Although complex, observations of CP have the potential to probe the physical structure and properties of active galactic nuclei ({AGNs}) deep into their cores.

Almost all measurements of extragalactic circular polarization have so far been made in AGNs, {for example, as revealed in} BL Lac objects \citep{gab03}, and blazars \citep{mys15}, that is, in sources for which strong AGN activity is clearly occurring and well-known.  For AGNs with jets, the circularly polarized signal is associated with the compact core \citep[e.g.][]{hom03}. {Rarely, CP can additionally be seen in inner bright jets; for example the MOHAVE VLBI survey at 15 GHz has found CP in 15\% of its 133 AGNs of which only are few examples of CP were detected in jets \citep{hom06}.}  The signal is often time-variable \citep{all03} although stable systems have also been observed {both in $m_C$ \citep[e.g.][over time periods of up to 5.7 years]{mys17} as well as in the sign of $m_C$ \citep[e.g.][over a time period of 20 years]{hom01}.}

Prior to the CHANG-ES {(Continuum Halos in Nearby Galaxies -- an EVLA Survey)} program \citep{irw12a}\footnote{{This survey targets 35 nearby edge-on galaxies at L-band (1.5 GHz) and C-band (6.0 GHz) in full polarization.  See \citet{irw12a} for details.}},
the only nearby systems showing CP were M~81 \citep{bru01,bru06} and Sgr A* in the Milky Way itself \citep{bow02,bow03,mun12}\footnote{The radio jet in NGC~1275 also shows strong $m_C$ \citep{hom04}, though at a distance of $\sim$ 70 Mpc, we do not classify it as `nearby'.}.  The discovery of strong CP, at a level of $\approx\,2\%$ in a Virgo Cluster CHANG-ES galaxy, NGC~4845, however \citep{irw15} has opened up the possibility of exploring new {\it nearby AGNs} in detail, especially AGNs that are embedded in other emission.

NGC~4845 was observed serendipitously about one year after a tidal disruption event ({TDE}) was detected in the hard X-ray regime; this event set the outburst time. The fact that CHANG-ES observations were broad-band yielding {\it in-band} spectral indices, and because multiple observations in different arrays were obtained, we were able to fit a simple AGN jet model with helical $B$ fields to this source.  Key to these measurements was the observation of steep negative spectral indices for Stokes $V$, $\alpha_V$,  which are predicted for Faraday conversion \citep[][their Eqn.~E11]{irw15} in the absence of complicated structural inhomogeneities \citep{jon77}. CP was detected at 1.6 GHz (hereafter {L-band}) near the turn-over to optical thickness, but not at 5 GHz (hereafter {C-band}) consistent with a steep $\alpha_V$.  The CP has now been confirmed through follow-up VLBI observations \citep{per17} at about the same percentage level.

We have now done a thorough search through all 35 CHANG-ES galaxies at L-band in our highest resolution ($\approx 3$ arcsec) B-configuration data, and here report on the detection of 4 more sources showing believable CP.  Together with NGC~4845, there are now 5 CHANG-ES galaxies that show CP in their cores.
Since optical spectroscopy surveys miss approximately half of the AGN population simply due to extinction through the host galaxy \citep{gou09}, radio detection of CP is an additional tool for identifying AGNs, especially low-luminosity AGN ({LLAGN}) \citep{pta01}, provided sensitive enough data are obtained.
Since {\it linear} polarization may be completely undetectable in compact cores due to depolarization, indeed CP may be the {\it only} way of detecting low luminosity radio AGNs with certainty when {they are embedded in nuclear star forming regions}.

The identification of AGN is also important for the far-infrared (FIR) - radio continuum relation. For example, \citet{won16} report that AGNs contribute significantly to this relation at 1.4 GHz; we also discuss the AGN affect on this relation in \cite{li16}.
Note that the detection of CP in a galaxy core is strong evidence that an AGN is present, but an AGN may still be present without necessarily showing CP. 
We will report, more generally, on AGNs in the CHANG-ES sample in a future paper.

In Sect.~\ref{sec:obs_data} we describe our observations and data reductions, Sect.~\ref{sec:results} outlines our results and presents the X-ray data. Sect.~\ref{sec:discussion} gives the discussion and Sect.~\ref{sec:conclusions} presents our conclusions.  In addition, Appendix~\ref{appendixA} shows the Faraday conversion development, progressing from the foundational work by \citet{bec02}.

\section{Observations and Data Reductions}
\label{sec:obs_data}

Observations were made with the  Karl G. Jansky Very Large Array (hereafter, the {JVLA}) using standard CHANG-ES data reduction and imaging procedures.
A complete description of the observing, data reduction and imaging strategies can be found in
\citet{irw12b} and
\citet{irw13}.  A brief description follows. 
\subsection{Observations}

B-configuration L-band observations were centerd at 1.575 GHz with a bandwidth extending from
1.247 to 1.503 and from 1.647 to 1.903 GHz
for a total of 512 MHz.  The central gap was set to avoid known interference.  
The L-band bandwidth contained 32 spectral windows, each with 64 channels for a total of 2,048 channels; however, Hanning smoothing was applied so that the effective channel width was doubled.

The observations were carried out within a `scheduling block' ({SB}),
in the standard fashion which included a single scan on the primary gain and phase
calibrator (hereafter, the primary calibrator) of known flux density and source structure
as well as a single scan on a polarization leakage calibrator which is known to have
negligible linear polarization {(OQ~208 was used for all galaxies except N~660 and N~891 which used 3C~84)}. The polarization calibrator (which was the primary gain and phase calibrator)
is used only for determination of the linear polarization (LP) and is not relevant for the CP {(but see comments in Sect.~\ref{sec:discussion})} which we focus on here (see below).  Observations on the source were flanked by observations of a secondary gain and phase
calibrator (hereafter, the secondary calibrator) which was close to the source on the sky.
Other galaxies were included in the SB and observations of any given galaxy were interspersed throughout the SB so that good uv 
distribution would result. The sources, their calibrators and observing dates are listed in Table~\ref{table:observations}.  {Information about these galaxies can be found in \citet{irw12a} and their distances are in \citet{wie15}.  We provide details, distances, etc.  for the CP detected galaxies in  Sects.~\ref{sec:N660} through \ref{sec:N4845}.}

{
\begin{deluxetable}{lccccc}
\hspace*{-10cm}
\tabletypesize{\scriptsize}
\renewcommand{\arraystretch}{0.9}
\tablecaption{B-configuration L-band\tablenotemark{a} Observing Data\label{table:observations}}
\tablewidth{0pt}
\tablehead{
  \colhead{Galaxy}  & \colhead{Obs. Date\tablenotemark{b}} & \colhead{RA{\tablenotemark{c}}} & \colhead{DEC{\tablenotemark{c}}}
  & \colhead{Primary Cal.\tablenotemark{d}}
& \colhead{Second. Cal.\tablenotemark{e}}  \\ 
}
\startdata
NGC~660    &  2012-06-24 & 01h43m02.40s & +13d38m42.2s &  3C~48  & J0204+1514  \\
NGC~891    &  2012-06-24 & 02h22m33.41s & +42d20m56.9s &  3C~48  & J0230+4032  \\
NGC~2613   &  2011-03-21 & 08h33m22.84s & -22d58m25.2s  &  3C~286 & J0856-2610  \\
NGC~2683   &  2012-06-16 & 08h52m41.35s & +33d25m18.5s   &  3C~286 & J0837+2454  \\
NGC~2820   &  2012-06-24 & 09h21m45.58s & +64d15m28.6s  &  3C~286 & J0921+6215  \\
NGC~2992   &  2011-03-21 & 09h45m42.00s & -14d19m35.0s  &  3C~286 & J0943-0819  \\
NGC~3003   &  2012-06-16 & 09h48m36.05s & +33d25m17.4s &  3C~286 & J0956+2515  \\
NGC~3044   &  2011-03-21 & 09h53m40.88s & +01d34m46.7s  &  3C~286 & J1007-0207  \\
NGC~3079   &  2012-06-23 & 10h01m57.80s & +55d40m47.3s &  3C~286 & J1035+5628  \\
NGC~3432   &  2012-06-24 & 10h52m31.13s & +36d37m07.6s   &  3C~286 & J1130+3815  \\
NGC~3448   &  2012-06-23 & 10h54m39.24s & +54d18m18.8s  &  3C~286 & J1035+5628  \\
NGC~3556   &  2012-06-23 & 11h11m30.97s & +55d40m26.8s &  3C~286 & J1035+5628  \\
NGC~3628   &  2012-07-30 & 11h20m17.01s & +13d35m22.9s &  3C~286 & J1120+1420  \\
NGC~3735   &  2012-06-24 & 11h35m57.30s & +70d32m08.1s  &  3C~286 & J1313+6735  \\
NGC~3877   &  2012-06-17 & 11h46m07.80s & +47d29m41.2s  &  3C~286 & J1219+4829  \\
NGC~4013   &  2012-08-11 & 11h58m31.38s & +43d56m47.7s &  3C~286 & J1146+3958  \\
NGC~4096   &  2012-08-11 & 12h06m01.13s & +47d28m42.4s  &  3C~286 & J1146+3958  \\
NGC~4157   &  2012-07-04 & 12h11m04.37s & +50d29m04.8s   &   3C~286 & J1219+4829  \\
NGC~4192   &  2012-07-30 & 12h13m48.29s & +14d54m01.2s &  3C~286 & J1254+1141  \\
NGC~4217   &  2012-08-11 & 12h15m50.90s & +47d05m30.4s  &  3C~286 & J1219+4829  \\
NGC~4244   &  2012-06-09 & 12h17m29.66s & +37d48m25.6s &  3C~286 & J1227+3635  \\
NGC~4302   &  2012-07-29 & 12h21m42.48s & +14d35m53.9s  &  3C~286 & J1254+1141  \\
NGC~4388   &  2012-07-29 & 12h25m46.75s & +12d39m43.5s &  3C~286 & J1254+1141  \\
NGC~4438$^f$& 2012-07-29 &  12h27m45.59s & +13d00m31.8s &  3C~286 & J1254+1141  \\
NGC~4565   &  2012-06-03 & 12h36m20.78s & +25d59m15.6s &  3C~286 & J1221+2813  \\
NGC~4594   &  2011-03-17 &  12h39m59.43s & -11d37m23.0s  &  3C~286 & J1248-1959  \\
NGC~4631   &  2012-06-03 & 12h42m08.01s & +32d32m29.4s  &  3C~286 & J1221+2813  \\
NGC~4666   &  2012-06-10 & 12h45m08.59s & -00d27m42.8s  &   3C~286 & J1246-0730  \\
NGC~4845   &  2012-06-11 & 12h58m01.19s & +01d34m33.0s &  3C~286 & J1407+2827  \\
NGC~5084   &  2011-03-17 & 13h20m16.92s & -21d49m39.3s   &  3C~286 & J0204+1514  \\
NGC~5297   &  2012-06-10 &  13h46m23.68s & +43d52m20.5s  &  3C~286 & J1327+4326  \\
NGC~5775   &  2011-04-05 & 14h53m58.00s & +03d32m40.1s &  3C~286 & J1445+0958  \\
NGC~5792   &  2011-04-05 &  14h58m22.71s & -01d05m27.9s  &  3C~286 & J1505+0326  \\
NGC~5907   &  2011-03-08 & 15h15m53.77s & +56d19m43.6s &  3C~286 & J1438+6211  \\
UGC~10288  &  2011-04-05 & 16h14m24.80s & -00d12m27.1s &  3C~286 & J1557-0001  \\
\tableline
\enddata
\tablenotetext{a}{The central frequency for all observations and resulting maps was 1.575 GHz, unless otherwise indicated.}
\tablenotetext{b}{Observing dates, designated Year-Month-Day (UT).}
\tablenotetext{c}{Center of galaxy, from the NASA Extragalactic Database ({NED}); the pointing center and field center were set to these values.}
\tablenotetext{d}{Primary gain and phase calibrator.}
\tablenotetext{e}{Secondary gain and phase calibrator.}
\tablenotetext{f}{Only the upper half of the band produced reliable results for NGC~4438, hence the central frequency for maps of this galaxy was 1.775 GHz.}
\end{deluxetable}}

\subsection{Signal Detection at the JVLA}

The JVLA uses circularly polarized right ($R$) and left ($L$) handed feeds and
a correlated output from any two antennas $RR$, $LL$ (the `parallel-hands') and
$RL$, $LR$ (the `cross-hands') are measured for each baseline.  These are related to the Stokes parameters via
\begin{eqnarray}
  I & = & (RR + LL)/2\\
  Q & = & (RL + LR)/2\\
  U & = & (RL - LR)/2i\\
  V & = & (RR - LL)/2\label{eqn:V}
\end{eqnarray}
where it is understood that these signals are averages over some integration time (10s for CHANG-ES).
The absolute scale is set by the primary calibrator.  Linear polarization and angle of a linearly polarized signal are then,
respectively,
\begin{eqnarray}
  P_{lin}&=&\sqrt{Q^2+U^2 - \sigma_{Q,U}^2}\label{eqn:lin_pol}\\
  \chi&=& \left(1/2\right)arctan\left(U/Q\right)
    \end{eqnarray}
where $\sigma_{Q,U}$ is the rms noise of the $Q$ and $U$ maps and does an approximate correction for
Ricean bias \citep{sim85,vai06,eve01}\footnote{Some authors increase $\sigma_{q,U}$ by a factor of $1.4$ which we have not done in this zeroth order approximation.  For a more sophisticated analysis, see \citet{mul17}.}.

The JVLA follows the IEEE standard such that a positive $V$ signal corresponds to a counter-clockwise rotation {of the E vector} when viewing the signal coming towards us and negative V corresponds to clockwise rotation\footnote{\tt https://library.nrao.edu/public/memos/evla/EVLAM\_195.pdf}.

\subsection{CHANG-ES Standard Calibrations and Imaging}
\label{sec:standard}

Data were reduced using the Common Astronomy software Applications ({CASA}) package\footnote{{\tt http://casa.nrao.edu}, using Version 4.7.2 (r39762) for $V$ or earlier versions for $I$. 
 }
\citep{mcm07}. {Calibration steps are standard for wide band data are are more thoroughly described in \citet{irw13}.
Calibration of RR and LL is based on a known model for the primary gain and phase calibrator \citep{per13}, usually 3C~286 (Table~\ref{table:observations}). RR and LL are calibrated separately.}  {Corrections were applied for antenna positions, delays, bandpasses,
and gain and phase as a function of time, the latter via the secondary gain and phase calibrators.} The data were Hanning smoothed and
flagged manually and several iterations of the calibration steps were carried out when additional flagging
occurred (and so throughout). 
 In each determination of
a new correction table, any previous tables were applied on the fly.  Corrections were applied to all calibrators as well. 


In addition, the cross-hands were also calibrated and imaged to obtain polarization and polarization angle images, but since linear polarization is not the focus of this paper, we omit those details in our description.  See \citet{irw12b} and
\citet{irw13} for further calibration details.

Imaging in Stokes $I$ was carried out using CASA's {\it clean} task, utilizing the multi-scale multi-frequency (ms-mfs) algorithm \citep{rau11}, including a wide field option \citep{cor08b}
with 128 w-projection planes
and Briggs robust = 0 uv weighting \citep{bri95}. During cleaning, an in-band spectral index, $\alpha$, is fitted, but no spectral index curvature term is included.
Self-calibration \citep[e.g.][]{pea84} was also attempted for the total intensity images but results were only kept if the dynamic range improved.
Total intensity 
images were then corrected for the primary beam ({PB}). 

\subsection{Special Considerations for Circular Polarization}
\label{sec:special}

A Stokes $V$ signal can be either positive or negative and the measurements are technically challenging.  This is because the signal is intrinsically weak and possibly variable (Sect.~\ref{sec:introduction}) and also because one must measure a difference (rather than the average) of signals (Eqn.~\ref{eqn:V}); those correlated signals (RR or LL) are sensitive to weather disturbances and radio frequency interference (RFI).
In addition, 
circularly polarized feeds are not the best choice for detection of circularly polarized signals because of possible instrumental effects that can carry through to create an apparent signal; linearly polarized feeds, such as used at the Australia Telescope Compact Array (ATCA) have advantages for observing CP \cite[e.g.][]{ray00,osu13}. Nevertheless, CP can still be measured with sufficient care; CP in M~81, Sgr A*, and CHANG-ES NGC~4845 (Sect.~\ref{sec:introduction}) were measured at the Very Large Array.  JVLA polarization characteristics are also known to be stable over timescales of months with polarization dynamic ranges of 10$^3$ possible \citep{sau13}.

Recently \citet{mys17} have looked carefully at how to correct for instrumental effects using circular feeds, finding typical uncertainties of $\approx\,0.1\%$ in $m_C$ using the Effelsberg 100 m telescope. 
However, 
CHANG-ES does not have sufficient calibrator coverage to replicate this approach.

{ A `gain transfer' procedure has also been described in \citet{hom99}, \citet{hom01}, and \cite{hom06}.  In this approach, a large number of sources are averaged and smoothed in time (RR and LL separately) to produce a smoothed set of gains that can be applied to all sources.  The calibration is carried out  iteratively so that outliers can be omitted prior to the final gain calibration.  This approach can work well with a large number of sources that are observed together, for example 40 sources in 48 hours as described in \citet{hom01}.
  By contrast, each CHANG-ES B-array L-band observation was 2 consecutive hours in duration with the entire sample observed over the course of
  17 months. Thus, individual calibration for each galaxy was required.

Therefore, of necessity, we have calibrated our sample in the standard fashion (Sect.~\ref{sec:standard}), but have attempted to incorporate
possible uncertainties into our error estimates (e.g. possible CP in the secondary calibrators) as described below.  Our approach is similar to that adopted by \citet[][their Sect. 3.3]{ray00}}.

Errors can be separated into {\it direction-independent errors} and {\it direction-dependent errors}.  The former relate to calibration errors prior to imaging and the latter relate to errors which enter into the imaging process \citep{bha08}.

{Direction-independent errors in $V$ include a possible calibration error due to the fact that the parallel hand leakage terms remain uncalibrated in our sample.  Such errors have a maximum value of \citep{hom99},
\begin{equation}
\frac{V}{I} = \frac{1}{\sqrt{N_s N_a \left(N_a-1\right)}}\left(\sqrt{2\left(N_a - 1\right)}\,m_L\,D + D^2  \right)
  \end{equation}
where $N_s$ is the number of scans that are separated in parallactic angle, $N_a$ is the number of antennas, $m_L$ is the fractional linear polarization, and $D$ is the fractional leakage term.  Here $N_a\,=\,27$.  For our detections (Sect.~\ref{sec:detections}), the parallactic angle coverage spans typically 109 degrees (minimum of 90 degrees and maximum of 160 degrees) in 6 different scans; thus we take $N_s\,=\,6$, or at least 15 degrees separating scans.  In no case do we detect any linear polarization at the location of the $V$ signal. Thus we assume an upper limit to  $m_L$ as the rms noise level of Q and U (which are the same) divided by the total intensity at the location of the $V$ signal.  Finally, we do not have a direct measurement of $D$, but have referred to a variety of sources that specify its value at the JVLA to be a few percent.  To overestimate this error, we take $D\,=\,0.05$.  For our detections, then, the largest error in $V/I$ as a result of neglecting possible leakage in the parallel hands is 0.004\%.  As this is significantly smaller than the other errors discussed below, it will not be considered further.
}

{A remaining possible source of calibration error is if the calibrators themselves might be circularly polarized. We therefore imaged the (calibrated) primary and secondary calibrators (Table~\ref{table:observations}) for each source and included a measurement of the residual calibrator signal into the uncertainty analysis.  We then take the uncertainty in $V/I$ to be,}

\begin{equation}
\sigma_{V/I} = \sqrt{{\sigma_{gal}}^2 + {\sigma_{prim}}^2 + {\sigma_{sec}}^2 }\label{eqn:uncertainty}
  \end{equation}
where
\begin{eqnarray}
\sigma_{gal} &= {\Delta V_{gal}}/I_{gal}\label{eqn:deltaVoverI}\\
\sigma_{prim} &= {\Delta V_{prim}}/I_{prim}\\
\sigma_{sec} &= {\Delta V_{sec}}/I_{sec}\\
\end{eqnarray}
In the above, the subscripts, $gal$, $prim$, and $sec$ refer to the galaxy, the primary calibrator, and the secondary calibrator, respectively.    

For the galaxy, $\Delta V_{gal}$ is the rms noise of the $V$ map.  For the point source calibrators, $\Delta V_{prim}$ and $\Delta V_{sec}$ are the rms values\footnote{{Note that we are using $\Delta$ for rms noise values of any image, whereas we are using $\sigma$ to denote the uncertainty in the ratio, $V/I$.}} 
of a small residual signal at the map center which, in all cases, was larger than the rms noise of the $V$ map as a whole.  The residual calibrator signal occurred and was measured over the region in which $I$ had contiguously positive values (essentially within the region in which the synthesized beam falls to zero). 
In the event that a residual $V$ is present in the calibrator, this error measurement should take such a signal into account. 
The galaxy signal was considered to be real only if $|m_C|\,>\,\sigma_{V/I}$.  The absolute value is used simply because CP can be positive or negative as indicated above.

{This approach is meant to be conservative so that we do not mistake calibration errors for real signals.  It is possible that real CP signals could still be present at lower levels but the observations would have to have been designed differently to detect them with confidence.}

{As for direction-dependent, or post-calibration errors, we first note that}
all of our sources were placed at the pointing center and at the imaging field center and each real source is very small ($\ltabouteq 20$ arcsec, Sect.~\ref{sec:results}).  The full-width at half-maximum (FWHM) of the PB at L-band is 25.8 arcmin \citep{per16} so corrections for the PB were not necessary for $V$.

Direction-dependent errors which have not been taken into account in our standard imaging are those due to the azimuthally asymmetric primary beam and its rotation on the sky, and the well-known beam squint due to the fact that R and L feeds point to slightly different positions on the sky \citep{bha08}. These affect off-center sources rather than those at the field center. Beam squint, in particular, can introduce a false $V$ signal, though not at the center \citep{bri03}.  Nevertheless, we made a variety of tests using the {\it awproject} algorithm in CASA\footnote{Casa Version 4.7.2-REL (r39762) was used.} which corrects for these effects, for two of our real signals.  In each case, the central signal remains although several minor off-center signals do not.


A remaining potential source of error relates to the application of self-calibration when amplitude is allowed to vary.
For the 5 sources showing CP, 3 of them included amplitude self-calibration.  For these, we either imaged the non-self-calibrated data or did a self-calibration that derives gains from the average of R and L hands ({\it gaintype='T'} in CASA {\it gaincal}). Again, the $V$ signal remained.

An interesting byproduct of this exercise is that we have 33 measurements of $V/I$ for 3C~286 itself.  We find
$(V/I)_{1.575~{\rm GHz}}\,=\,+0.086\,\pm\,0.072\%$ which is marginally positive. 
Thus, either there is a residual instrumental offset in our JVLA observations at this level, or 3C~286 has a marginal positive CP at L-band.  

Finally, we make no attempt to correct for a possible thermal contribution to Stokes $I$. The thermal contribution at L-band is typically $\approx\,10$\%  for galaxies as a whole \citep[e.g.][]{con92} which is considerably less than our uncertainties in $V/I$ (Table~\ref{table:detections}).  We also argue in this paper for the Faraday conversion interpretation which requires that non-thermal particles dominate in our detected AGNs.

{
\begin{deluxetable}{lcccc}
\hspace*{-10cm}
\tabletypesize{\scriptsize}
\renewcommand{\arraystretch}{0.9}
\tablecaption{L-band Upper Limits\tablenotemark{a}\label{table:nondetections}}
\tablewidth{0pt}
\tablehead{
  \colhead{Galaxy}  & 
   \colhead{Beam size\tablenotemark{b}} & \colhead{RA\tablenotemark{c}} & \colhead{DEC\tablenotemark{c}}
  & \colhead{$(V/I)_L$  \tablenotemark{d}}  \\
     &   (arcsec, arcsec, deg)& (\%)
}
\startdata
NGC~891    &  3.09, 2.88, 53.46 & 02h22m33.23s & +42d20m58.3s  &  $<0.36$  \\
NGC~2613   & 5.28, 2.97, 0.20 & 08h33m22.79s & -22d58m25.2s  & $<8.77$ \\
NGC~2683   & 3.02, 2.95, 55.86 &  08h52m41.31s & +33d25m19.0s   &$<1.40$\\
NGC~2820   & 3.23, 3.17, 52.76  &  09h21m45.87s & +64d15m27.5s  & $<1.87$ \\
NGC~2992   & 4.87, 3.57, 16.39 & 09h45m41.93s & -14d19m36.0s  &  $<0.09$  \\
NGC~3003   &  3.04, 2.96, 70.84& 09h48m36.69s & +33d25m17.9s &  $<19.0$  \\ 
NGC~3044   & 3.53, 3.34, 36.97 & 09h53m40.88s & +01d34m46.7s  &$<0.45$  \\
NGC~3432   &  3.20, 3.12 82.82 & 10h52m30.96s & +36d37m08.6s   & $<11.1$ \\ 
NGC~3448   &  3.11, 2.95, 62.42 &  10h54m39.19s & +54d18m20.8s  & $<0.58$ \\
NGC~3556   & 3.08, 2.96, 56.36 & 11h11m30.43s & +55d40m26.7s &  $<0.88$  \\ 
NGC~3735   & 3.24, 3.11, 33.77& 11h35m57.20s & +70d32m07.6s  & $<0.77$  \\
NGC~3877   & 3.01, 2.83, 19.18 & 11h46m07.70s & +47d29m39.7s  & $<0.87$  \\
NGC~4013   & 3.01, 2.90, -84.20 & 11h58m31.38s & +43d56m51.1s &$<0.34$  \\
NGC~4096   & 3.06, 2.94, -84.88& 12h06m01.25s & +47d28m40.6s  & $<13.0$  \\ 
NGC~4157$^e$   & 3.01, 2.83, 26.37& 12h11m04.37s & +50d29m04.8s   & $<4.97$ \\ 
NGC~4192   & 3.21, 3.07, -7.49 & 12h13m48.29s & +14d54m02.1s & $<0.37$  \\
NGC~4217   & 3.05, 2.92, -84.62 & 12h15m50.90s & +47d05m29.4s  &  $<0.64$  \\
NGC~4244$^e$   & 3.09, 3.00, 44.97 & 12h17m29.66s & +37d48m25.6s & $<41.4$ \\
NGC~4302   & 3.48, 3.10, -8.53& 12h21m42.31s & +14d35m52.4s  & $<1.18$ \\
NGC~4438   & 3.32, 2.91, -6.28 & 12h27m45.52s & +13d00m33.2s &  $<0.06$ \\
NGC~4565   & 3.22, 2.96, 41.96 &12h36m20.78s & +25d59m15.6s & $<0.95$  \\
NGC~4594   & 4,36, 3,25, -14,03 &  12h39m59.43s & -11d37m23.0s  &  $<0.18$ \\
NGC~4631   & 3.36, 3.03, 62.61 &  12h42m07.87s & +32d32m34.9s  &$<0.54$  \\
NGC~4666   & 3.75, 3.43, 34.26 & 12h45m08.62s & -00d27m42.8s  & $<0.31$ \\
NGC~5084   & 5.63, 2,95, -11,79& 13h20m16.85s & -21d49m38.3s   & $<0.36$  \\
NGC~5297   & 3.13, 2.99, 52.75&  13h46m23.67s & +43d52m20.4s  &$<10.8$  \\
NGC~5775   & 3.57, 3.40, 54.80 & 14h53m57.34s & +03d32m43.8s & $<0.69$  \\  
NGC~5792$^e$   & 3.89, 3.42, 48.63 & 14h58m22.71s & -01d05m27.9s  & $<0.40$  \\ 
NGC~5907   & 3.35, 2.79, -4.64 & 15h15m53.47s & +56d19m43.6s & $<4.14$  \\
UGC~10288  & 3.67, 3.50, 57.18 & 16h14m24.79s & -00d12m27.2s & $<11.0$  \\
CHANG-ES A$^f$ & 3.67, 3.50, 57.18& 16h14m23.28s & -00d12m11.6s $<0.15$   \\
\tableline
\enddata
\tablenotetext{a}{Upper limits were calculated using Eqn.~\ref{eqn:uncertainty} (see Sect.~\ref{sec:special}). Positions are as in Table~\ref{table:observations} unless otherwise noted.}
\tablenotetext{b}{Synthesized beam parameters: Major axis, Minor axis, Position Angle.}
\tablenotetext{c}{Position of the peak in total intensity, $I$, which was closest to the NED position of Table~\ref{table:observations}, unless otherwise indicated.
  This is also where $V$ was measured.}
\tablenotetext{d}{Circular Polarization upper limits expressed as a fraction of the total intensity at the same position.}
\tablenotetext{e}{Measurements were made at the NED center.  In these cases, total intensity emission is seen from the disk but little emission was seen at the center.}
\tablenotetext{f}{CHANG-ES A is the bright double-lobed radio source behind the disk of UGC~10288 \citep{irw13}.  This measurement was made at the center of the background source.}
\end{deluxetable}}

{
\begin{deluxetable}{lcccccc}
\hspace*{-10cm}
\tabletypesize{\scriptsize}
\renewcommand{\arraystretch}{0.9}
\tablecaption{L-band Detections\label{table:detections}}
\tablewidth{0pt}
\tablehead{
  \colhead{Galaxy}  & \colhead{$V_{{max}_L}$\tablenotemark{a}}& \colhead{$\Delta\,V_{{gal}_L}$\tablenotemark{b}}& \colhead{$I_{{max}_L}$\tablenotemark{a}}
  & \colhead{Beam size\tablenotemark{c}}
  & \colhead{$(V_{max}/I_{max})_L$  \tablenotemark{d}} & \colhead{$(S_V/S_I)_L$\tablenotemark{e}} \\
    &  ($\mu$Jy beam$^{-1}$)   &  ($\mu$Jy beam$^{-1}$) & (mJy beam$^{-1}$) & (arcsec, arcsec, deg)&(\%) & (\%)
}
\startdata
NGC~660    & +715.0    & 50.0 & 245.0 & 3.39, 3.27, 44.43 & $+0.29\,\pm\,0.07$ & $+0.31\,\pm\,0.07$\\
NGC~3079   & -221.9    & 19.0 & 122.7 & 3.14, 3.00, 58.44 & $-0.18 \,\pm\,0.15$ & $-0.17\,\pm\,0.2$  \\
NGC~3628   & -198.7    & 13.5 & 80.0  & 3.21, 3.13, 3.73 & $-0.25\,\pm\,0.08$ & $-0.27\,\pm\,0.08$\\
NGC~4388   & -583.0    & 25.0 & 25.2  & 3.57, 3.22, -2.09 & $-2.31\,\pm\,0.10$ & $-2.6\,\pm\,0.3$\\
~~~~~~~~~Jet& -145.8   & 25.0 & 4.48  & 3.57, 3.22, -2.09 & $-3.26\,\pm\,0.10$ & $-3.4\,\pm\,0.3$  \\
NGC~4845   & +3700    & 24.0 & 209.3 &3.55, 3.34, 27.73 & $+1.77\,\pm\,0.11$ & $+2.3\,\pm\,0.2$\\
\tableline
\enddata
\tablenotetext{a}{$V_{max}$ and $I_{max}$ are measured at the same position. They are also at the same NED center as given in Table~\ref{table:observations} to within 0.7 arcsec, except for: a) NGC~660 whose peaks are located 3 arcsec to the north-west of the NED center, b) NGC~3628 which are 3 arcsec south of the NED center, and c) the NGC~4388 Jet which is measured at a local peak at
RA = 12$^{\rm h}$ 25$^{\rm m}$ 46$\rasec$85, DEC = 12$^\circ$ 39$\decmin$ 50$\decsec$5 and is marked with a blue `+' in Fig.~\ref{fig:allgalaxies}.}
\tablenotetext{b}{$V$ map noise of the image (see Sect.~\ref{sec:special}).}
\tablenotetext{c}{Synthesized beam parameters: Major axis, Minor axis, Position Angle.}
\tablenotetext{d}{CP percentage with uncertaintly calculated from Eqn.~\ref{eqn:uncertainty}.}
\tablenotetext{e}{CP percentage using the ratio of flux densities (rather than specific intensities) measured over a half-power-beam-width centered at $|V_{max}|$. Uncertainties are propagated from the previous column with increases, if required, for variations that result from adjusting the central position by approximately a cell size.}  
\end{deluxetable}}











\section{Results}
\label{sec:results}

\subsection{L-band Non-Detections of Stokes $V$}
\label{sec:nondetections}

Table~\ref{table:nondetections} lists $m_C\,=\,V/I$ upper limits for the galaxies which were not detected at B-configuration L-band along with the beam sizes and measurement positions in the event that future observations may wish to compare with these data.  In addition, we include CHANG-ES A which is a bright source almost directly behind UGC~10288 \citep{irw13}.

{Measurements were made at the total intensity, $I$, peak that was closest to the NED center of the galaxy as given in Table~\ref{table:observations} except for 3 sources that had no clear $I$ peak near the core (see Table footnotes) in which case we measured at the NED center itself.  An example is NGC~4244 which has a very high upper limit to $V/I$ because of its low $I$ value.
  With the exception of only two sources (next paragraph), no $V$ signal was measured at all.
 Consequently, $V/I$ was simply taken to be the the rms noise value in $V$ divided by $I$.

The two exceptions were NGC~4594 and NGC~5084 which showed weak $V$ signals at 4.1 times and 3.0 times their respective rms $V$ noise values. Once uncertainties in the calibration were factored in, however, the resulting $V$ signals were not considered to be significant.
  }


{Of this list, at the corresponding location, only a single galaxy (NGC~4438) shows any {\it linear} polarization, so at B-configuration L-band, there is essentially no detectable linear polarization for the CHANG-ES galaxies}.  This appears to be because the signal is weak in these small ($\approx\,3$ arcsec) beams and also, when observing edge-on galaxies, there is significant depolarization along the line of sight. That CP can be detected for some galaxies (next section) whereas linear polarization usually cannot, is an argument for including CP measurements in future observations of highly compact sources.

\subsection{L-band Detections of Stokes $V$}
\label{sec:detections}

Relevant data for the 5 galaxies that exceed the limits specified above are given in Table~\ref{table:detections}. In each case the $V$ signal was measured at its peak value which is at or near the galaxy center (see Notes to the table). The level at which $V$ has been detected ranges from 12$\Delta V$ to 154$\Delta V$ ($\Delta V$ being the rms noise). Notice that the signals are both positive and negative, as one would expect from natural signals rather than systematic offsets.

The weakest signal, and most marginal detection is NGC~3079; its $V$ signal is measured at the 12$\Delta V$ level. 
Two galaxies, however, show very strong $V/I$ in their cores, of order 2\%, namely NGC~4388 and NGC~4845.
{For NGC~4845, there is a small difference between the ratio of the peak $V/I$ and the ratio of flux densities, $S_V/S_I$ because the total intensity distribution is slightly more peaked than the CP distribution at the center.
Recall that the signal is only considered to be real if $V/I$ exceeds the errors as defined in Eqn.~\ref{eqn:uncertainty}; hence signal-to-noise alone for $V$ is not taken to definitively determine whether a CP signal is significant.}

$V$ and $I$ images are shown in Fig.~\ref{fig:allgalaxies}. Although CP has been reported previously in NGC~4845 \citep{irw15}, we present the $V$ image for the first time here.

{An interesting result is that all sources except for NGC~3079 show $V$ emission that is resolved and shows structure, sometimes rather complex. The most striking case is NGC~4388 which shows CP in the direction of its northern jet}.  As indicated in Sec.~\ref{sec:introduction}, few examples of CP associated with jets exist in the literature.  We will discuss each galaxy further in Sects.~\ref{sec:N660} to \ref{sec:N4845}.

For each detection, we also checked for CP at C-band {reduced in the same way as described for L-band}.  The results, with other relevant data, are given in Table~\ref{table:cband}.  In no case do we detect any CP at the higher frequency. This is an important result because it implies that the $V$ spectral index is very steep, consistent with the interpretation of Faraday conversion as discussed further in Sec.~\ref{sec:predictions}.

{We also do not detect any linear polarization at the location of the CP signal, again consistent with the Faraday conversion mechanism that  we discuss at length in this paper though Faraday depolarization is also likely at play \citep{irw15}.}

{
\begin{deluxetable}{lcccc}
\hspace*{-10cm}
\tabletypesize{\scriptsize}
\renewcommand{\arraystretch}{0.9}
\tablecaption{C-band Upper Limits for Detected Sources\tablenotemark{a}\label{table:cband}}
\tablewidth{0pt}
\tablehead{
  \colhead{Galaxy}  & \colhead{Obs. Date\tablenotemark{b}}&   \colhead{Beam size\tablenotemark{c}}&\colhead{$\Delta\,V_{{gal}_C}$\tablenotemark{d}}&\colhead{$(V/I)_C$\tablenotemark{e}}  \\
    & &  (arcsec, arcsec, deg) &($\mu$Jy beam$^{-1}$)   &  (\%) 
}
\startdata
NGC~660    &2012-01-27 &   3.15, 2.64, -55.72&  14.2  &$<0.015$  \\
           &2012-01-28 &         &         &           \\
NGC~3079   &2012-02-15 &  2.68, 2.59, -88.97 & 3.35  &$<0.081$    \\
NGC~3628   &2012-02-29 & 3.68, 2.74, -78.98& 3.10  &$<0.022$  \\
NGC~4388$^f$   &2012-04-08 & 2.79, 2.72, -13.15& 3.20  &$<0.039$ \\
NGC~4845   &2012-02-23& 3.15, 2.76, 35.56& 13.6  &$<0.052$  \\
           &2012-02-25&       &         &          \\
\tableline
\enddata
\tablenotetext{a}{Positions are the same as the L-band detection positions. The central frequency is 6.00 GHz.}
\tablenotetext{b}{Observing Dates, designated Year-Month-Day (UT).  If two dates are given, the observing was split into two sessions.}
\tablenotetext{c}{Synthesized beam parameters: Major axis, Minor axis, Position Angle.}
\tablenotetext{d}{$V$ map noise level (see Sect.~\ref{sec:special}).}
\tablenotetext{e}{Circular Polarization percentage upper limits calculated from Eqn.~\ref{eqn:uncertainty}.}
\tablenotetext{f}{Values for the jet are the same.}
\end{deluxetable}}

\subsection{Spectral Index Measurements}
\label{sec:alpha_measurements}

We distinguish between {\it band-to-band} spectral indices and {\it in-band} spectral indices.  Our spectral index results are listed in Table~\ref{table:spectral_indices}.

{
\begin{deluxetable}{lcccccc}
\hspace*{-10cm}
\tabletypesize{\scriptsize}
\renewcommand{\arraystretch}{0.9}
\tablecaption{Spectral Indices\tablenotemark{a}\label{table:spectral_indices}}
\tablewidth{0pt}
\tablehead{
  \colhead{Galaxy}  & \colhead{$\alpha_{I_L}$\tablenotemark{b}}& \colhead{$\alpha_{V_L}$\tablenotemark{c}} & \colhead{$\alpha_{I_C}$\tablenotemark{d}}& \colhead{$p$\tablenotemark{e}} &
  \colhead{$\Delta\,T$\tablenotemark{f}} &\colhead{$\alpha_V(L-C)$\tablenotemark{g}} \\
    & & & & & (days)&\\
}
\startdata
NGC~660    & $+1.23$ & $-1.4$ & $-0.314$ & 1.63  &149, 148 &  $<-2.9$ \\
NGC~3079   & $-0.41$ & --- & $+0.295$ & --- &129 & $<-3.1$  \\
NGC~3628   & $-0.31$ & $ +2.7$ & $-0.675$ & 2.35 &152 & $<-3.1$\\
NGC~4388   & $-0.94$ & $-2.7$ & $-0.818$& 2.64 &112 & $<-3.9$\\
~~~~~~~~~Jet&$-1.05$ &  ---   & $-0.805$& 2.61 & 112 & $<-2.9$        \\
NGC~4845$^h$   & $+0.81$ & $-3.4$ & $-0.493$& 1.99 &109, 107 & $<-4.2$\\   
\tableline
\enddata
\tablenotetext{a}{Positions are the same as in Table~\ref{table:detections}. Values are in-band spectral indices except for the last column.  A horizontal line means that there was insufficient signal-to-noise to provide a reliable value.  Estimated errors in $\alpha{_I{_L}}$, $\alpha{_V{_L}}$ and $\alpha{_I{_C}}$ are $\approx$ 5\%, 20\% and 1\%, respectively. }
\tablenotetext{b}{Total intensity spectral index at L-band measured from the spectral index maps which were formed as described in \citet{wie15}.  Values are the average over the L-band half-power beam width centered at $|V_{max}|$.}
\tablenotetext{c}{Stokes $V$  spectral index at L-band. Signals which are positive (see last column of Table~\ref{table:detections}) were measured in the same way as for $\alpha_{I_L}$. Signals which are negative were fitted as described in Sect.~\ref{sec:alpha_measurements} and illustrated in Fig.~\ref{fig:alpha}. Dashes indicate that there is insufficient signal in the subbands for a reliable result.}
\tablenotetext{d}{Total intensity spectral index at C-band measured in the same way as $\alpha_{I_L}$.}
\tablenotetext{e}{Energy spectral index of relativistic particles ($N(E)\,\propto\,E^{-p}$) assuming that the C-band spectral index (except for NGC~3079) is optically thin, i.e. $p\,=\,1\,-\,2\alpha_{I_C}$.}
\tablenotetext{f}{Duration between the earlier C-band observations and later L-band observations. Galaxies which had split observations (Table~\ref{table:cband}) have two entries.}
\tablenotetext{g}{Upper limits to the L-band to C-band spectral indices using Eqn.~\ref{eqn:bandtoband}.}
\tablenotetext{h}{From \citet{irw15}.}
\end{deluxetable}}


Band-to-band spectral indices, $\alpha_V(L-C)$, are measured between the center of L-band and the center of C-band.  We do not detect CP at C-band, so we can only provide upper limits to $\alpha_V(L-C)$ using,
\begin{equation}\label{eqn:bandtoband}
  \frac{|V_L|}{\Delta V_{{gal}_C}} \,=\,\left(\frac{\nu_L}{\nu_C}\right)^{\alpha_V (L-C)}
    \end{equation}
where $V_L$ is the detected L-band signal (Table~\ref{table:detections}) and $\Delta V_{{gal}_C}$ is the rms noise at C-band (Table~\ref{table:spectral_indices}) {since there is no emission or any sign of residual signals above the rms noise at C-band}.  These upper limits are also estimates because of the possibility that the CP is variable, hence we note the dates and number of days that have elapsed between L-band and C-band observations. For example, for NGC~4845, variability of order 20\% is observed in L-band quantities over a time period of 164 days and the flux density in Stokes, $I$ at C-band varies by 18\% over only 67 days \citep{irw15}.

In-band spectral indices are generated when spectral fitting is carried out during the mapping and cleaning process.  The wide bands used in CHANG-ES permits such measurements, providing a slope for the center of any given band.  The default cut-off is 5$\sigma$ for the formation of spectral index maps.

However, due to a CASA
limitation\footnote{In-band spectral indices are not computed when the signal is negative.}, we must determine in-band spectral indices manually for any $V$ signal that is negative.  Consequently, for such cases, we have split the band into 4 equal frequency sections and imaged each section separately. The resulting maps are then smoothed to the resolution of the map at the lowest frequency end of the band.  We then examine whether the signal at each of the 4 frequencies has a signal-to-noise (S/N) $>\,5\sigma$. If not, we conclude that $\alpha_{V_L}$ cannot be reliably obtained.  If so, we then determine $\alpha_V$ via a curve fit using the
Levenberg-Marquardt algorithm \citep{lev44, mar63}.

\begin{figure*}
\rotatebox{0}{\scalebox{0.5} 
             {\includegraphics[height=8.0truein,width=7.5truein]{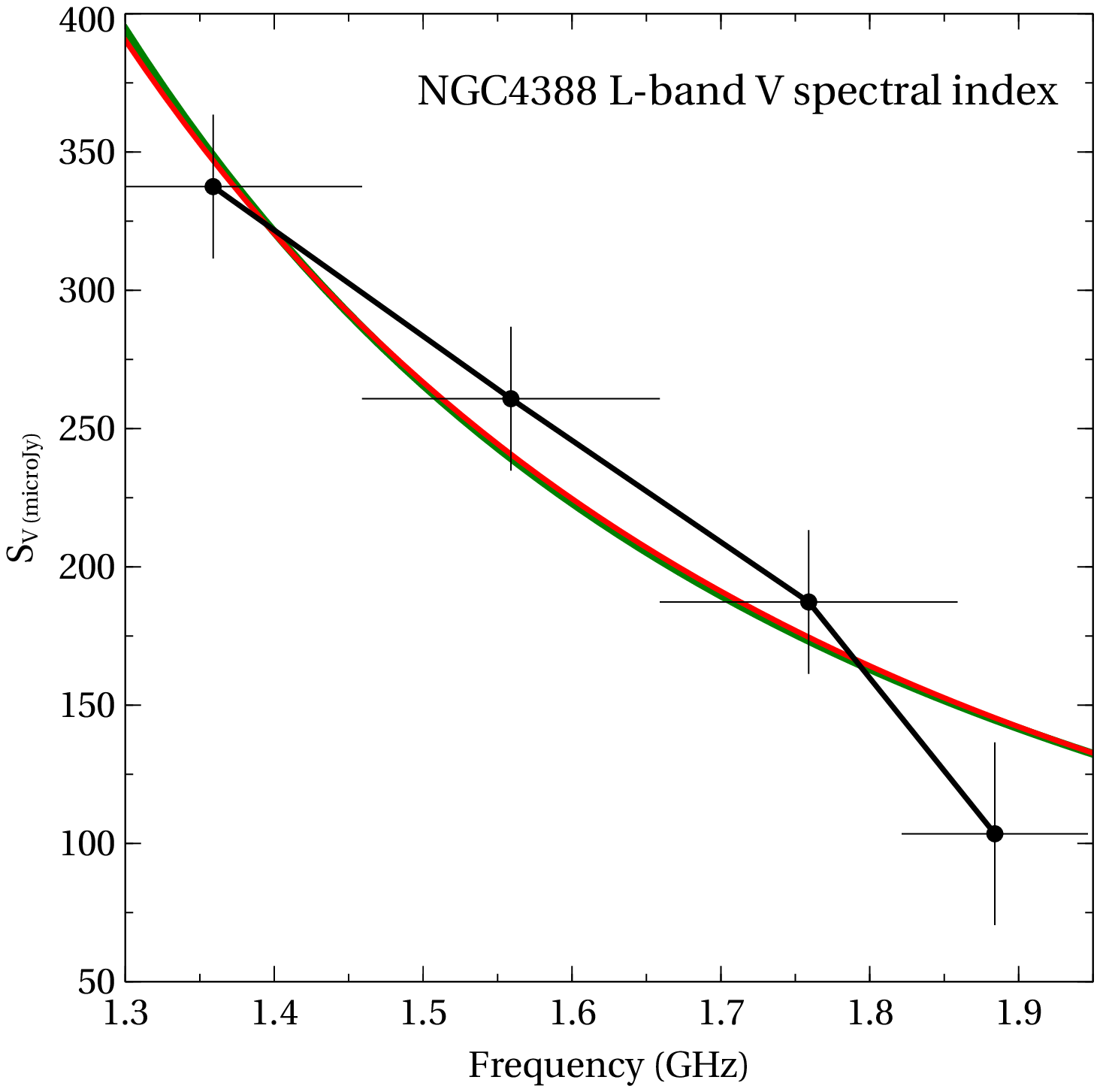}}}\\  
\caption{L-band $V$ signal (inverted to be positive) as a function of frequency for NGC~4388, where the band has been broken into 4 frequency sections to see the spectral dependence.  Red and green curves (virtually indistinguishable) represent best fits of the form,
  $V=a\nu^{\alpha_V}$ where $a=784.8$ and $\alpha_V=-2.7$ (red, reduced $\chi^2\,=\,1.3$), and
  $V=C_1\nu^{-1} + C_2\nu^{-3}$ where $C_1=54.7$, and $C_2=773.8$ (green, reduced $\chi^2\,=\,1.4$), for $\nu$ in GHz and $S_V$ in $\mu$Jy. For the last point, only 5/8 of the frequency band was useable.
}
\label{fig:alpha}
\end{figure*}

An example is shown in Fig.~\ref{fig:alpha} for NGC~4388, where negative values (which simply describe the direction of the CP) have been inverted to positive in order to show the plot in the standard fashion.  The red curve (indistinguishable from the green)
 corresponds to $V=a\nu^{\alpha_{V_L}}$ and the green curve will be discussed in Sect.~\ref{sec:predictions}. 


In Table~\ref{table:spectral_indices}, we also list the spectral indices for Stokes $I$ in each band.  This allows us to check whether CP is occurring in an optically thick or optically thin regime and also provides an estimate of $p$ in the optically thin regime, should either be the case.

\subsection{The Faraday Conversion Interpretation}\label{sec:predictions}

Faraday conversion has not yet been established for all sources and source complexities may also be present.  For example, optical depth effects can lead to a variety of spectral indices, as pointed out by \citet{jon77}, and variable (including inverted) Stokes $V$ spectral indices have been observed in M~81 \citep{bru06} and Sgr A* \citep{bow03}.  Nevertheless, given that this interpretation has achieved the most success at explaining the observations to date (Sect.~\ref{sec:introduction}), we have developed and expanded the analysis of \citet{bec02} in order to provide analytical solutions for a variety of relatively straightforward cases.

\subsubsection{Faraday Conversion Predictions for $p\,=\,2$}
\label{sec:p=2}

In \citet{irw15}, Appendix E, we determined the frequency dependence of Stokes $V$ for the case in which the power law spectral index of relativistic electrons, $p\,=\,2$ (where $N(E)\,=\,N_0\,E^{-p}$) for the energy distribution of electrons. This choice of $p$ was adopted because the observed radio continuum total intensity {\it in-band} spectral index, $\alpha_{I_C}\,\approx\,-0.5$ in the C-band observations of NGC~4845.  This implies that, at C-band, the core of NGC~4845 is optically thin in which case $\alpha_I\,=\,(1-p)/2$.  At L-band, however, the core is transitioning to being optically thick which is where CP is observed.

{In our homogeneous source with the Faraday rotation coefficient small compared to the conversion and absorption coefficients \citep{irw15}, but nevertheless optically thin,}
the flux density of Stokes $V$, $S_V$, has a frequency dependence of,
\begin{eqnarray}
S_V&=& \frac{C_1}{\nu}\,+\,\frac{C_2}{\nu^3}.~~~~~~~(p\,=\,2)\label{eqn:faraday}
\end{eqnarray}
Here the $C_n$ values are functions of various physical parameters. 
The first term is an emission term and the second is the conversion term.  The conversion term requires that relativistic particles dominate.

{In fact, as found in \citet{jon77}, \citet{jon88}, and \citet{irw15}, circular polarization of the magnitude detected here requires a lower electron energy cut-off, $\gamma_0\,\approx\,\gamma_e$, where $\gamma_e$ is the Lorentz factor of an electron radiating at the peak radio frequency.  However, this does not explain the lack of observed linear polarization \citep{jon77} which must be due to a surrounding depolarizing sheath, in our view.

In \citet{jon77a}, the percentage circular polarization emerging from an optically thin boundary is calculated under the conditions mentioned above.  Generally the amplitude and frequency dependence is similar to what we observe and predict (Eqn.~\ref{eqn:faraday}), but there can be sudden cut-offs at high frequency probably due to a lack of Faraday rotation.  In a standard jet model \citep{jon88, irw15} the rotation of a helical magnetic field also serves to rotate the plane of polarization.
}
  
The observed L-band {\it in-band} spectral indices for NGC~4845, $\alpha_V$, range from 2.2 to 3.4, depending on JVLA configuration (i.e. spatial resolution) clearly showing that, in this interpretation, the medium is dominated by relativistic particles. \citet{osu13} also found an average $\nu^{-3}$ dependence for the quasar, PKS~B2126-158.  That is, the steep spectral dependence provides the evidence that Faraday conversion is the dominant mechanism and, for NGC~4845, also explains why CP is observed at L-band but not at C-band.


We now have 
  a tool for probing the physical properties deep within AGN that are otherwise inaccessible in total intensity observations.
In the case of NGC~4845 {\citep[see][]{irw15}}, our AGN jet model provided additional constraints on the physical parameters.  To explain the magnitude of the observed CP and its spectrum at 1.5 GHz, our estimated values were $B\,=\,0.04$ Gauss, {the relativistic electron density}, $n_{er}\,=\,100$ cm$^{-3}$, $L\,=\,10^{17}$ cm and initial linearly polarized Stokes $U_0\,=\,60$ mJy.  {The electron spectrum is a power law with lower cut-off near $\gamma_e$ for electrons radiating in the 1.5 GHz band.} {As stated above}, an additional depolarizing Faraday screen was also required to reduce the observed {\it linear} polarization to observable values; in an edge-on galaxy, such reduction is to be expected.


\subsubsection{Faraday Conversion Predictions for a general $p$}\label{sec:pne2}
 
We have now developed the CP predictions for the general case in which the power law spectral index for relativistic electrons, $p$, can have a range of values and provide straightforward analytical results in Appendix~\ref{appendixA}.

The emission term (Eqn.~\ref{emission}) gives a $\nu^{-1}$ dependence as in Eqn.~\ref{eqn:faraday} for $p=2$.  This is, however, flatter than  observed and so we concentrate mainly on the steeper conversion term.

The conversion term (Eqns.~\ref{convp<2practical}, \ref{convp=2practical}, and \ref{convp>2practical} for $p<2$, $p=2$, and $p>2$, respectively) reveals the steeper frequency dependence of Faraday conversion and shows the physical dependences explicitly.  For the $p=2$ case, we also now include a weaker logarithmic term that was omitted from \cite{irw15} Appendix E.

A general consequence is that CP should have a frequency dependence of $\nu^{-(2+p/2)}$ which goes as $\nu^{-3}$ when $p=2$ as found earlier.
For $p<2$ and the second term in Eqn.~\ref{convp<2practical} negligible, then the frequency dependence would be $\nu^{-(2+p/2)}$ as above.  If $p>2$ and the second term in Eqn.~\ref{convp>2practical} negligible, then the frequency dependence would be $\nu^{-3}$, also as found for $p=2$.  

The terms in square brackets, however, can modulate this behaviour, depending on the magnitude
of the term, $x\equiv 0.0028{\gamma_0}^2 B_\perp/\nu_9$, where $B_\perp$ is the perpendicular magnetic field,  $\nu_9$ (GHz) is the observing frequency,
and  $\gamma_0$ is the Lorentz factor of the lower energy cutoff of the relativistic electrons.
If $x^{1-p/2}\,{\rm is}\,\orderof\left(1\right)$\footnote{Of order one.}
 for example, then the magnitude of CP can decline and eventually reverse sign.
  We provide possible examples for which this modulating term may not be neglible in Sects.~\ref{sec:N660_conversion} and \ref{sec:N3628_conversion}.
If the observed frequency dependence follows the description given in the previous paragraph, though, then this modulating term is likely small.

\subsection{Reconciliation of Theory with Observations}
\label{sec:reconciliation}

The important and most clear-cut parameter, in the context of the CHANG-ES data, involves the frequency dependences given in Table~\ref{table:spectral_indices}.  A clear result is that the {\it band-to-band} spectral indices in every case are very steep, with an average of $\alpha_V(L-C)<-3.4$ for the galaxy cores. To explain this spectral dependence with an emission term (Eqn.~\ref{emission}) would require an electron energy index of $p=6.8$ (Eqn.~\ref{emission}), an unphysically steep value and unsupported by the data in the table.  We conclude that Faraday conversion can account for these steep spectral indices, although there are some peculiarities as discussed below.

Four sources provide Stokes $V$ {\it in-band} spectral index measurements, ${\alpha_V}_L$.  Of these, the two strongest sources, NGC~4388 (CP of -2.6\%) and NGC~4845 (CP of +2.3\%) also show very steep in-band spectral indices: ${\alpha_V}_L$ of -2.7 and -3.4, respectively (typical uncertainties of $\approx 20\%$, Table~\ref{table:detections}). These two are the most clear cut cases. NGC~4845 has been extensively examined in the context of a jet model in \cite{irw15} and need not be examined further here.

In the next 3 subsections, we consider each of the remaining 3 galaxies, namely the strong CP source, NGC~4388, as well as the weaker sources, NGC~660 and NGC~3628 (CP of 0.31\% and 0.27\%, respectively).

\subsubsection{Faraday Conversion in NGC~4388}\label{sec:N4388_conversion}

For NGC~4388, the conversion falls into the category, $p>2$ (Eqn.~\ref{convp>2practical}).  Although we do not have information on all of the physical parameters that enter into a CP calculation, we can ask what reasonable values might fit the observations.  For example, 
at L-band, no {\it linear} polarization is measured at the core, precluding an estimate of a lower limit for $U_o$.  In principle, we could make an 
 estimate of the amount of L-band linear polarization that is to be Faraday converted by observing the linear polarization at C-band along with its spectral index, and then extrapolating to L-band.  Unfortunately, the C-band linear polarization at the nucleus is also very weak, likely due to significant Faraday depolarization; with a degree of polarization at the nucleus of only $\approx\,0.2\%$ in C-band \citep{dam16} and a peak C-band value of 9.0 mJy beam$^{-1}$ (for the same spatial resolution), this corresponds to only 18 $\mu$Jy beam$^{-1}$ at C-band -- much less than the observed CP at L-band.  In other words, the observed C-band linear polarization extrapolated to L-band is far too weak to put meaningful constraints on $U_0$. Since Stokes $V$ at L-band is  $|{S_V}_L|\,=\,355 ~\mu$Jy, there must originally have been at least this amount of linearly polarized signal for conversion.


We do, however, have a limit on the line-of-sight distance through the source, $L$, since VLBI observations by \citet{gir09} provide a source size of 6 mas = 0.48 pc.  We also know $p=2.64$, $\nu_9=1.5$, and can adopt a reasonable value of $\theta=\pi/4$.
We do not know the other quantities in Eqn.~\ref{convp>2practical}
but, as an example, we can adopt values similar to those from the model of NGC~4845 \citep{irw15}. Thus for
$n_e\,=\,200$ cm$^{-3}$, $B\,=\,0.04$ Gauss, and $\gamma_o=100$, we find a (not unique) solution of
$|U_o/{S_V}_L|\,= 14\%$ conversion. Note that
no reasonable combination of values produces the $B=67~\mu$Gauss that is measured for the nucleus by \citet{dam16}, so it is clear that the CP is probing deep into the core of the AGN where the magnetic field is much higher.

CP is observed in the jet of NGC~4388 also and is remarkably strong, even stronger than at the core ($m_C\,=\,3.4\%$, Table~\ref{table:detections}).  We have measured jet values at the position of the peak $V$ intensity in the jet, marked with a blue `plus' in Fig.~\ref{fig:allgalaxies}.  We note, however that, again, the band-to-band spectral index is steep, as expected for Faraday conversion and spectral indices where measurable (Table~\ref{table:spectral_indices}), are similar to values in the core. Further discussion is in Sect.~\ref{sec:N4388}.

\subsubsection{Faraday Conversion in NGC~660}\label{sec:N660_conversion}

NGC~660 is an interesting case because its in-band spectral index (${\alpha_V}_L={-1.4}$) is much flatter than its band-to-band spectral index ($-2.9$), indicating curvature in the Stokes $V$ spectrum (Table~\ref{table:spectral_indices}).  For this galaxy, $p=1.63$ so the relevant equation is Eqn.~\ref{convp<2practical}.  We have no other information to help determine its physical parameters but note that a flatter distribution of relativistic particles generally results in stronger CP, all else being equal.
On the other hand, the second term in the square brackets must {\it not} be negligible for this source since, if it were, then
${S_V}_{conv}\propto \nu^{-2.8}$ which is contrary to observations.   Thus  we should see ${S_V}_{conv}\propto {\nu_9}^{-2.8}[1\,-\,x^{0.19}]$, where
$x$ was defined in Sect.~\ref{sec:pne2}.

Strictly speaking, $x$ cannot be equal to 1 or CP will disappear.  However, if this term is to modify the slope then,
to order of magnitude, we can let ${x}\approx 1$, or equivalently, $0.0028{\gamma_o}^2B\sin\theta=\nu_9$. For $\theta=\pi/4$, then ${\gamma_o}^2 B\approx 758$. If $\gamma_o=100$, we require $B \approx 0.08$ Gauss, for example.  Compared to the previous example, then, one can put tighter restrictions on the physical parameters when there are departures from the predicted frequency behaviour for conversion.  Note that $x$ increases with frequency within the band which will flatten the spectral index, as observed.


\subsubsection{Faraday Conversion in NGC~3628}\label{sec:N3628_conversion}

\begin{figure*}[h]
\rotatebox{0}{\scalebox{0.5} 
             {\includegraphics[height=7.0truein,width=6.5truein]{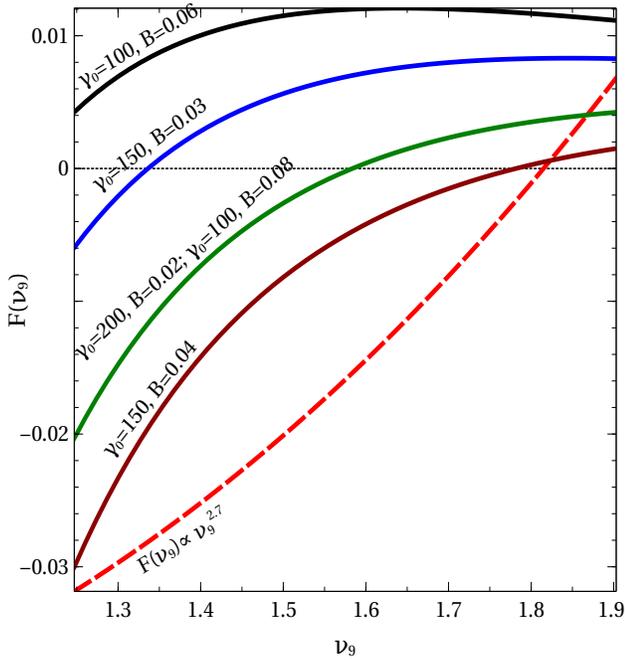}}}\\  
\caption{The function $F(\nu_9)$ (Eqn.~\ref{modulating}) as a function of frequency across L-band. The values of $\gamma_o$ and $B$ are marked on each of the solid curves (black, blue, green, and red).  For any of these curves, one could obtain an equivalent result if $\gamma_o$ is halved and $B$ is quadrupled, as illustrated for the green curve.  Note that $F(\nu_9)$ is actually {\it increasing} when  $x$ is 
$\orderof\left(1\right)$, in contrast to the normal steeply declining behaviour for Faraday conversion.  The red dashed curve plots the observed behaviour for NGC~3628 (with arbitrary scaling) as described in Sect.~\ref{sec:N3628_conversion}.
}
\label{fig:positive}
\end{figure*}

NGC~3628, like the other galaxies, has a steep band-to-band spectral index (Table~\ref{table:spectral_indices}), arguing for the Faraday conversion interpretation.  However, its Stokes $V$ in-band spectral index is positive (${\alpha_V}_L = + 2.7$) which is highly discordant in comparison to the declining spectral indices seen for the other sources.
It may be that absorption effects can turn over the spectral index, making it positive.  Indeed, the measurement of ${\alpha_V}_L$ is based on splitting a weak signal into 4 parts, as described in Sect.~\ref{sec:alpha_measurements} and is prone to errors.  Nevertheless, this is a good example to show how the spectrum can not only be flattened (as for NGC~660) but can actually become inverted.  In Fig.~\ref{fig:positive} we plot the term, $F(\nu_9)$, as given in Eqn.~\ref{modulating} which folds together the frequency dependent terms from Eqn.~\ref{convp>2practical}.  In this way, we can examine cases for which $x$ is 
$\orderof\left(1\right)$.

Here we see a series of curves for different values of $\gamma_o$ and $B$. The result is striking in that {\it increasing} values with frequency result, similar to what is observed for NGC~3628 (red dashed curve).  They are inverted with respect to the {\it steep negative} behaviour that is typically seen in Faraday conversion, a result that is solely due to the fact that
$x$ is 
$\orderof\left(1\right)$.
It is clear that the curvature of $F(\nu_9)$ does not match the fitted curve for the galaxy; however, given the uncertainties in this exercise, we take this as a good demonstration of the kind of spectral behaviour that could result.  If we now repeat the exercise carried out for NGC~660 by letting  ${x}\approx 1$, we would find similar values for $B$ and $\gamma_o$.

\subsection{X-ray Emission from the Galaxy Cores}
\label{sec:x-ray_cores}

\begin{deluxetable}{lcccccr}
\hspace*{-10cm}
\tabletypesize{\scriptsize}
\renewcommand{\arraystretch}{0.9}
\tablecaption{\label{table:xrays}Parameters of the XMM-Newton observations and parameters derived from the model fits to the core regions}
\tablewidth{0pt}
\tablehead{
  \colhead{Galaxy}
&  \colhead{ObsID}
  &  \colhead{Obs. Date}
  & $\Delta\,T$\tablenotemark{a}
&  \colhead{Foreground N$_H$\tablenotemark{b} }
&  \colhead{Internal N$_H$\tablenotemark{c}}
  & \colhead{Luminosity\tablenotemark{c}}\\
    \colhead{ }
&  \colhead{ }
    &  \colhead{ }
    &  \colhead{(days)}
&  \colhead{ (10$^{20}$\,cm$^{-2}$) }
&  \colhead{ (10$^{22}$\,cm$^{-2}$) }
  & \colhead{(10$^6$\,L$_{\odot}$) }\\
}
\startdata
NGC\,660        & 0093641001	& 2001-01-07	& 4186	        & 4.64 					& 	0.60$^{+0.64}_{-0.46}$		&	1.22$^{+2.83}_{-0.52}$		\\
\vspace{5pt}
		& 0671430101	& 2011-07-18 & 	342		&					&					&					\\
NGC\,3079       & 0110930201	& 2001-04-13	&  4089             & 0.89					&	--				&	2.30$^{+0.69}_{-0.51}$		\\
\vspace{5pt}
		& 0147760101	& 2003-10-14 &	3175		&					&					&					\\
\vspace{5pt}
NGC\,3628       & 0110980101	& 2000-11-27 &   4263                & 1.97 					&	0.33$\pm$0.04			&	1.89$^{+0.34}_{-0.29}$		\\
\vspace{5pt}
NGC\,4388       & 0110930701	& 2002-12-12 & 	3518        & 2.58 					&	26.45$^{+1.53}_{-1.47}$		&	599$^{+298}_{-191}$		\\
\vspace{5pt}
NGC\,4845       & 0658400601	& 2011-01-22 & 	507        & 1.44	  				&	6.93$\pm$0.10			&	1400$^{+102}_{-98}$		\\
\tableline
\enddata
\tablenotetext{a}{Duration between the earlier XMM observations and later B-array L-band observations.}
\tablenotetext{b}{Weighted average value after LAB (Leiden/Argentine/Bonn) Survey of Galactic {H}{I}
  \cite{kal05}.}
\tablenotetext{c}{{Luminosity in the 0.3 to 12 keV band,} derived from the model fits to the spectra. See text for details.}
\end{deluxetable}

\begin{figure*}
\rotatebox{-90}{\scalebox{0.48} 
             {\includegraphics[height=9.0truein,width=7.5truein]{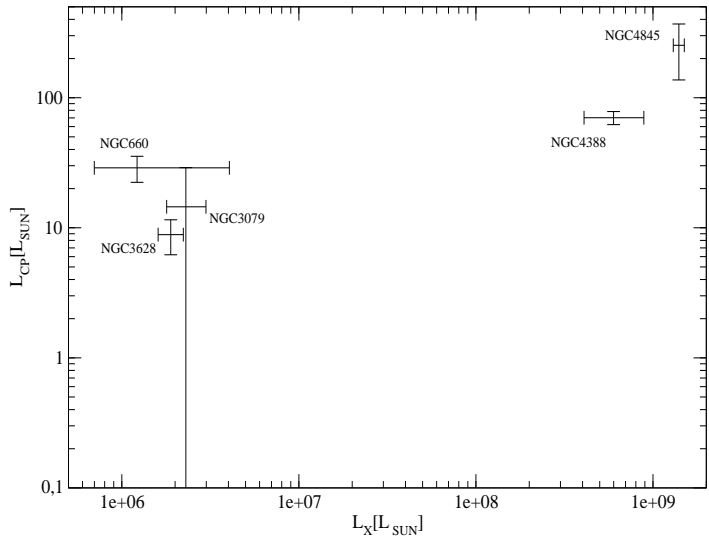}}}\\  
\caption{Luminosity of the circularly polarized signal (obtained from the flux density, $S_V$) against X-ray luminosity, both in units of the Solar luminosity.
}
\label{fig:xraygraph}
\end{figure*}

{{As has been shown in \citet{irw15} there is a close link between the radio and X-ray emission in NGC~4845.  The radio observations were obtained approximately one year after a tidal disruption event which was observed in the hard X-ray regime.  A declining light curve was observed in the radio and a model developed to quantitatively explain the observations as well as offer a possible link between the X-ray and radio regime.  This source showed strong CP, suggesting that it is important to measure the X-ray characteristics of our CP detections for current and subsequent analyses. All galaxies showing CP have been observed and detected in X-rays.  However, we have now re-analyzed the X-ray data in a consistent fashion.}

For the X-ray analysis of the core emission, XMM-Newton archive data were used. The data were processed using the SAS\footnote{Science Analysis System} 15.0.0 package \citep{gab04}
with standard reduction procedures. The tasks $epchain$ and $emchain$ helped to obtain event lists for two EPIC\footnote{European Photon Imaging Camera}-MOS cameras \citep{tur01} and the EPIC-pn camera \citep{str01}. 
The event lists were then carefully filtered for periods of intense background radiation by creating light curves of high-energy emission. With these light curves good time interval (GTI) tables 
were produced and used to remove the data when high count rates were observed.

Next, the spectral analysis was performed. The spectra of the central regions of all five galaxies were created using all three EPIC cameras. 
The position and sizes of the regions were chosen to both match the radio centre of a galaxy and to include the brightest X-ray emission. 
The background spectra were obtained using blank sky event lists \citep[see][]{car07}.
These were filtered using the same procedures as for the source event lists.
For each spectrum, response matrices and effective area files were produced.
Next, including these ancillary files, spectra from all three EPIC cameras and the corresponding background blank sky spectra were merged 
using the SAS task $epicspeccombine$ into a final background subtracted source spectrum. In the case of NGC\,660 and NGC\,3079, where two observations were available 
for each galaxy, the resulting spectra were created in the same way, as $epicspeccombine$ allows to merge also a larger number of spectra.
The spectra were then fitted using XSPEC~12 \citep{arn96}.

Because of the limited resolution of the XMM-Newton and consequently the sizes of the core regions used (between 25 and 61 arcseconds in diameter), 
one expects a contribution from the gaseous (thermal) component in the extracted spectra. Therefore, all spectra except one, were fitted with models that included both 
the gaseous and the central source components, represented by a {\it mekal} model and a  power-law model, respectively. {A mekal model is a model 
 of an emission spectrum from hot diffuse gas based on the calculations of Mewe and Kaastra \citep{mew85,kaa92}}.
The only spectrum that did not require a gaseous model component was that of NGC\,4845, most likely due to the very high brightness of the central source, which 
made the emission from the hot gas emission negligible.  
For all sources except NGC\,3079, an additional component of the model needed to be used to account for a high absorption in the core. For this galaxy, the contribution from the thermal component reached 24\% of the total emission, while for the other objects it was not higher than 6\% of the total X-ray core emission.  {\it It is important to note that the fluxes are measured in the range, 0.3 - 12 keV, uniformly for all galaxies.}

The fluxes of the power-law component derived directly from the model fits were then used to calculate the luminosities, which together with 
basic information on the used XMM-Newton datasets and absorption values (both foreground and fitted for the central regions) are presented in Table~\ref{table:xrays}. 

Fig.~\ref{fig:xraygraph} presents the relation between X-ray luminosities of the central sources and luminosities of the circular polarization component 
produced in the core regions of the galaxies. {As can be seen by the error bars, the relation is quite approximate but provides a baseline for possible future studies. As was indicated at the beginning of Sect.~\ref{sec:x-ray_cores}, there appears to be a close link between the X-ray data and CP.  Here we have ensured that all X-ray data were reduced in the same way so as to make a proper comparison. In Sect.~\ref{sec:summary_intercomparison} we discuss the important result of the internal absorption.   }
}

\section{Discussion}\label{sec:discussion}

In the next sections, we refer to Fig.~\ref{fig:allgalaxies} and discuss each of the galaxies in Table~\ref{table:detections} {and Table~\ref{table:xrays}}, followed by a discussion of the sample as a whole.

\subsection{NGC~660}\label{sec:N660}

NGC~660 \citep[D = 12.3 Mpc,][]{wie15} is a polar ring galaxy that has undergone a merger \citep{bra93,alt98}. Previous radio continuum data have been obtained by \citet{vand95,fil02} and \citet{fil04}.  Its nuclear type is HII/LINER\footnote{Low Ionization Nuclear Emission Region} \citep[NED,][]{irw12a}.  However, an activity type of Sy 2 has also been recorded \citep{ver06} and it is becoming clear that nuclear activity levels and classifications can change with time \citep[e.g.][]{lam15}.

\citet{arg15} found that NGC~660 has undergone a spectacular radio outburst which occurred some time between 2008
and 2010.7 (see also \cite{min13}) and our L-band data were taken after this outburst (June, 2012, Table~\ref{table:observations}). Their October 2013 high resolution EVN\footnote{European VLBI Network} image shows two new components, likely related to this outburst, one to the west and one to the north-east of the central bright source, suggesting jet-like outflow on scales of $\approx$ 1 pc.

The X-ray data of \citet{arg15}, aside from emission along the main NE-SW disk (see our Fig.~\ref{fig:allgalaxies}a plus NE to SW dashed line), appears to show an extension perpendicular to the disk on scales of  $\approx$ 2 arcsec (120 pc).  On a similar scale, we also see a SE `bulge' in the CP image contours, which we have emphasized with a NW to SE dashed line.  If these features are related, then it is likely that our SE bulge is also related to jet-like outflow (rather than a starburst) as suggested by \citet{arg15}.  However, such emission must be associated with a previous outflow event since highly superluminal motion would be required otherwise.  Our SE bulge/jet feature seen in CP is now one of the few sources known for which CP is seen in an extended structure; CP is usually only observed in a galaxy's core (Sect.~\ref{sec:introduction}).

{
  The two X-ray data points of the core (Table~\ref{table:xrays}) were taken before (first line) and after (second line) the putative radio outburst described by \citet{arg15}.  The first measurement (in 2001) was long before the radio outburst.  The second measurement (in 2011) was made between $\approx$ 1294  and 340 days after the outburst, given the uncertainty in the outburst window noted above.  Since the X-ray luminosities of the two measurements are not very different (within the error bars quoted) then if there had been an X-ray outburst associated with the radio outburst, it is likely that the X-ray outburst had already declined significantly by the time of the second measurement.}

As noted in Table~\ref{table:detections}, the total intensity as well as CP centers are offset to the NW of the optical center as given in NED.  A similar offset in the same direction has been noted by \citet{ste05} for the submm center.  The true center of the galaxy is likely at the radio and submm peaks. We find the radio peak to be at 
RA = 01$^{\rm h}$ 43$^{\rm m}$ 02$\rasec$33, DEC = +13$^\circ$ 38$\decmin$ 44$\decsec$7 to an accuracy of 0.3 arcsec in both coordinates at the observing date given in Table~\ref{table:observations}.

\subsection{NGC~3079}\label{sec:N3079}

NGC~3079 \citep[D = 20.6 Mpc,][]{wie15} has a unique radio structure in the form of two kpc-scale radio lobes on either side of the NW-SE major axis, the  (Fig.~\ref{fig:allgalaxies}b, colour image), first detected by \citet{deb77}. The AGN in this galaxy has been detected in VLBI, showing linear features related to a jet, luminous H$_2$O maser emission and a time variable component \cite[see summary in][]{kon05}. X-ray observations support the presence of an obscured AGN \citep{cec02, iyo01} and \citet{mid07} have detected variably brightening components in VLBI due to jet-cloud interactions. Recently, \citet{sha15} have found a rich interplay between NGC~3079 and its nearby companions.

This is our weakest CP signal (Fig.~\ref{fig:allgalaxies}b) and only the central core is represented.  We were unable to measure a reliable in-band spectral index in Stokes $V$.  However, the total intensity spectral index is negative at L-band and positive at C-band.  This sugests that at least 2 components must be present within our 3 arcsec (300 pc) beam; for example, one with a negative spectral index dominating at L-band and one with a positive spectral index becoming dominant at C-band. \citet{mid07} have mapped the multiple VLBI components in this source, showing that some have positive and some negative spectral indices which also vary with time.

\subsection{NGC~3628}\label{sec:N3628}

NGC~3628 \citep[D = 8.5 Mpc,][]{wie15} is a member of the Leo Triplet, and has an HII/LINER nuclear classification as listed in NED.  Due to its high inclination and prominent dust lane, however, its AGN is highly obscured.
 \citet{gou09} list it as an `optically unidentified' AGN which they identified via the high excitation emission line, [NeV]$\lambda14.32~\mu$m.  The AGN was previously identified, however, via the presence of a hard nuclear X-ray source \citep[][and Table~\ref{table:xrays}]{gon06,yaq95}.  \citet{flo06} also fit a diffuse X-ray spectrum in the central kpc of this galaxy.  \citet{don06} find a black hole mass of $2\,\times\,10^7$ M$_\odot$.  


Our observations show some evidence for a resolved source with two extensions emerging NE to SW from the nucleus (Fig.~\ref{fig:allgalaxies}c) and extending $\approx\,10$ arcsec ($\approx\,400$ pc).  The soft X-ray CHANDRA data also show emission in these directions at about the same scale and the hard X-ray emission shows the SW extension \citep[][their Fig. 8]{tsa12}. There is an enhancement in roughly this direction in the inner 4 arcsec of the galaxy as seen in the mid-IR \citep{asm14}.  Given the additional evidence for alignments in the NE-SW direction, our CP results suggest that a jet is in this direction.

\subsection{NGC~4388}\label{sec:N4388}

NGC~4388 \citep[D = 16.6 Mpc,][]{wie15} is in the Virgo Cluster
and harbours an outflow lobe, {i.e. the vertical jet that extends away from the plane to the north,} that is clearly seen in the colour total intensity image of Fig.~\ref{fig:allgalaxies}d. {Note that the plane of the galaxy is east-west so the jet is approximately perpendicular to the plane in projection.}
The total intensity and linear polarization of NGC~4388 have been reported by \citet{dam16} for C-band only, revealing new linearly polarized features and their interaction with the intergalactic medium (IGM). NGC~4388 is a hard X-ray source \citep[][{and Table~\ref{table:xrays}}]{kri15,iwa03} and has been detected in VLBI at 1.6 GHz \citep{gir09}.

This galaxy shows the strongest $m_C\,=\,V/I$ in the CHANG-ES sample and the core of this galaxy
shows the `classic' CP spectrum characteristic of Faraday conversion (Sect.~\ref{sec:N4388_conversion}). Since the linear polarization of the AGN is entirely depolarized at L-band and almost entirely depolarized at C-band, CP in NGC~4388 appears to be the {\it only} way to probe the physical properties of the AGN at these frequencies.  Known and adopted physical parameters produce results that are quite reasonable, given our development in Appendix~\ref{appendixA}.

{Remarkably, there is CP along the northern jet as well, as high as $\approx\,$3\% at the peak $V$ position {of the jet} that is marked with a blue `+' in Fig.~\ref{fig:allgalaxies}d.  This is probably the clearest case yet of observations of CP in a jet. At this position, there is $-146~ \mu$Jy beam$^{-1}$ in Stokes $V$, yet there is no B-configuration L-band linearly polarized emission at all, to an rms level of $12~\mu$Jy beam$^{-1}$, supporting the Faraday conversion interpretation.  At C-band, there is a strongly {\it linearly} polarized lobe centered at the position marked with a black `x' in our figure \citep[see also Fig. 3 of][]{dam16}. However, this emission falls off strongly towards the south such that there is only marginal linear polarization at the position of our blue `+'.  Thus, the C-band linear polarization is anti-correlated spatially with the CP at L-band.  Again, this supports the Faraday conversion interpretation.}

The emission is not strong enough in the jet to measure $V$ in-band spectral indices from our data.  However, given how extensively this galaxy has been observed by others, there may be additional constraints that can be applied to this region in the future. 
More sensitive observations with the application of the same kind of analysis as in Sect.~\ref{sec:N4388_conversion} may yield further details of the physical conditions in this jet.



\subsection{NGC~4845}\label{sec:N4845}

NGC~4845 (D = 17 Mpc) has been extensively studied in \citet{irw15} who discovered the CP and provided an initial development of the Faraday conversion model for the case, $p=2$ (now superseded by our Appendix~\ref{appendixA}).  In this paper,  a jet model was fit to the core which explained the observed variability in the multiple CHANG-ES observations, and also assisted with the CP interpretation.

{Our B-array L-band radio data were taken after a hard X-ray outburst detected by \citet{nik13} who interpreted the outburst as being due to the tidal disruption of a Jupiter-mass object by a black hole.  The peak of the X-ray curve has been well determined to be Jan. 22, 2011, 507 days prior to our B-array L-band observations.  We have now {re-analyzed the data corresponding to this X-ray peak  (Table~\ref{table:xrays}), ensuring 
that the analysis has been done in the same fashion for all of our CP-detected galaxies.}}

Follow-up VLBI observations have now confirmed the CP at the $\approx$ 2\% level \citep{per17}, in agreement with the CHANG-ES observations. The VLBI observations have resolved the CP, showing that it is elongated NW-SE with a size $\approx\,5$ mas (0.4 pc).  At the same angle in the NW direction, there is a feature at L-band that is possibly related to nuclear outflow.  Our observations are of much lower resolution (3.4 arcsec, or 280 pc), but there is a hint of this angle in our CP map as well, with contours that bulge out in the same direction.   We have drawn a NW-SE dashed line through this slight bulge in our $V$ image (Fig.~\ref{fig:allgalaxies}e, see below).  The fact that the level of CP is the same between VLBI and our CHANG-ES data suggests that most of the CHANG-ES-detected CP is concentrated into a small region associated with the AGN.  Indeed, this is consistent with our Faraday conversion parameters in the context of the jet model.

Fig.~\ref{fig:allgalaxies}e shows a map of the CP for NGC~4845 for the first time, since this map had not yet been shown in
\citet{irw15}.  Although most of the CP is concentrated in a small region at the core, we also see that it is resolved, similar to most of the other galaxies.
An interesting new result is that the CP occurs mainly in an elongation that extends from the north-east (NE) to the south-west (SW) at an angle that is tilted with respect to the total intensity. We have marked this disk by a NE-SW dashed line in Fig.~\ref{fig:allgalaxies}e and have labelled it: the `conversion disk'.  The conversion disk has an extent of $\approx~15$ arcsec (1.2 kpc).  This is almost as large as the total intensity disk \citep[1.8 kpc][]{irw15} which delineates the galaxy's major axis.

Although the conversion disk is tilted with respect to the galactic disk, it is perpendicular to the likely outflow direction as described above. {Consequently, we suggest that it is the conversion disk that is associated with the outflow alignment.  The {\it core size}, within which most of the {\it total intensity} flux is measured, is less than 41 pc \citep{irw15}.  Thus any accretion disk is no doubt much smaller than the conversion disk; possibly the inner `edge' of the conversion disk is actively feeding the AGN and aligning the outflow in this galaxy. Recently \citet{mck13} have developed models showing various alignments of accretion disks with black hole spins or with galaxy disks, depending on disk thickness and distance from the black hole.
{An example of a misaligned accretion disk in a nearby LLAGN is NGC~4258 \citep{her05}.}

  The tilted conversion disk of NGC~4845 also requires that a kpc-scale magnetic field be present that is thus tilted with respect to the major axis.
  For such a disk to exist requires that relativistic electrons not only dominate (as they likely do in the inner total intensity disk as well) but the physical parameters in this disk are such that conversion can occur. 
  }

To our knowledge, this is the first time that {such a feature has been seen in any galaxy and we have coined the term, `conversion disk', to reflect our interpretation that the CP mechanism is conversion, as well as the fact that this disk-like feature has a large extent.}



\begin{figure*}
  \centering
  \begin{subfigure}{}
        \centering
        \includegraphics*[width=0.48\linewidth]{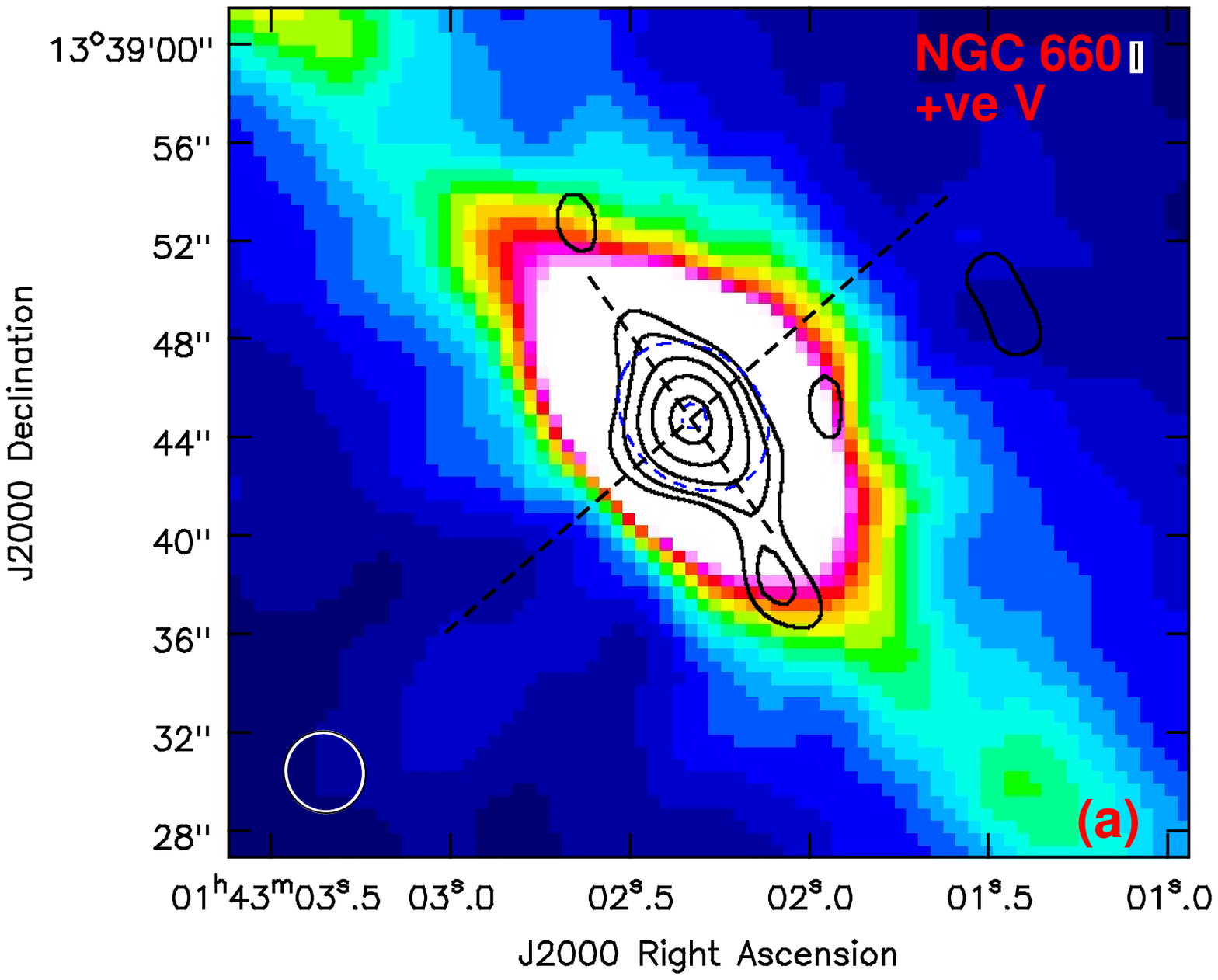}
        \includegraphics*[width=0.48\linewidth]{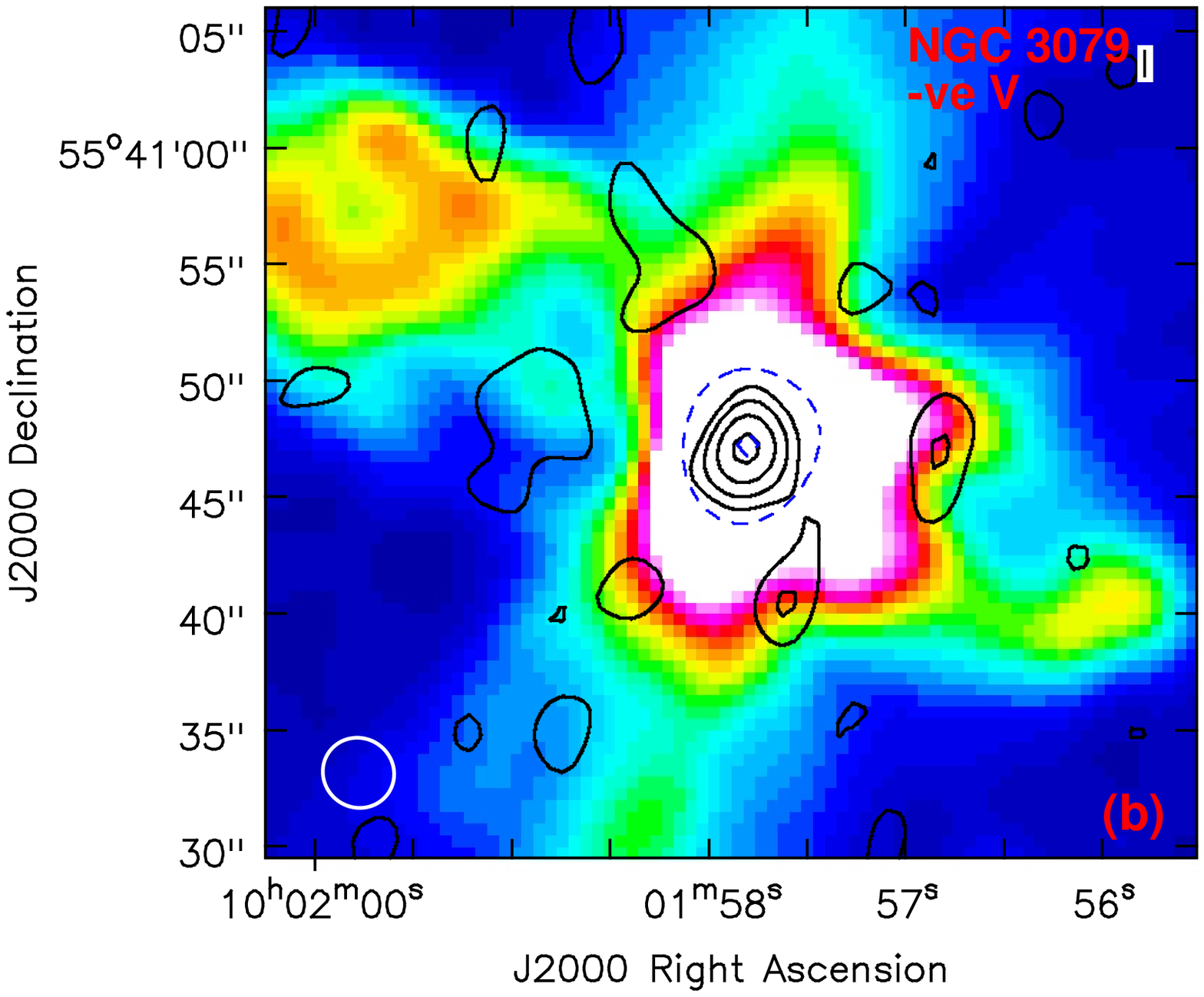}
    \end{subfigure} %
    \begin{subfigure}{}
        \centering
        \includegraphics*[width=0.45\linewidth]{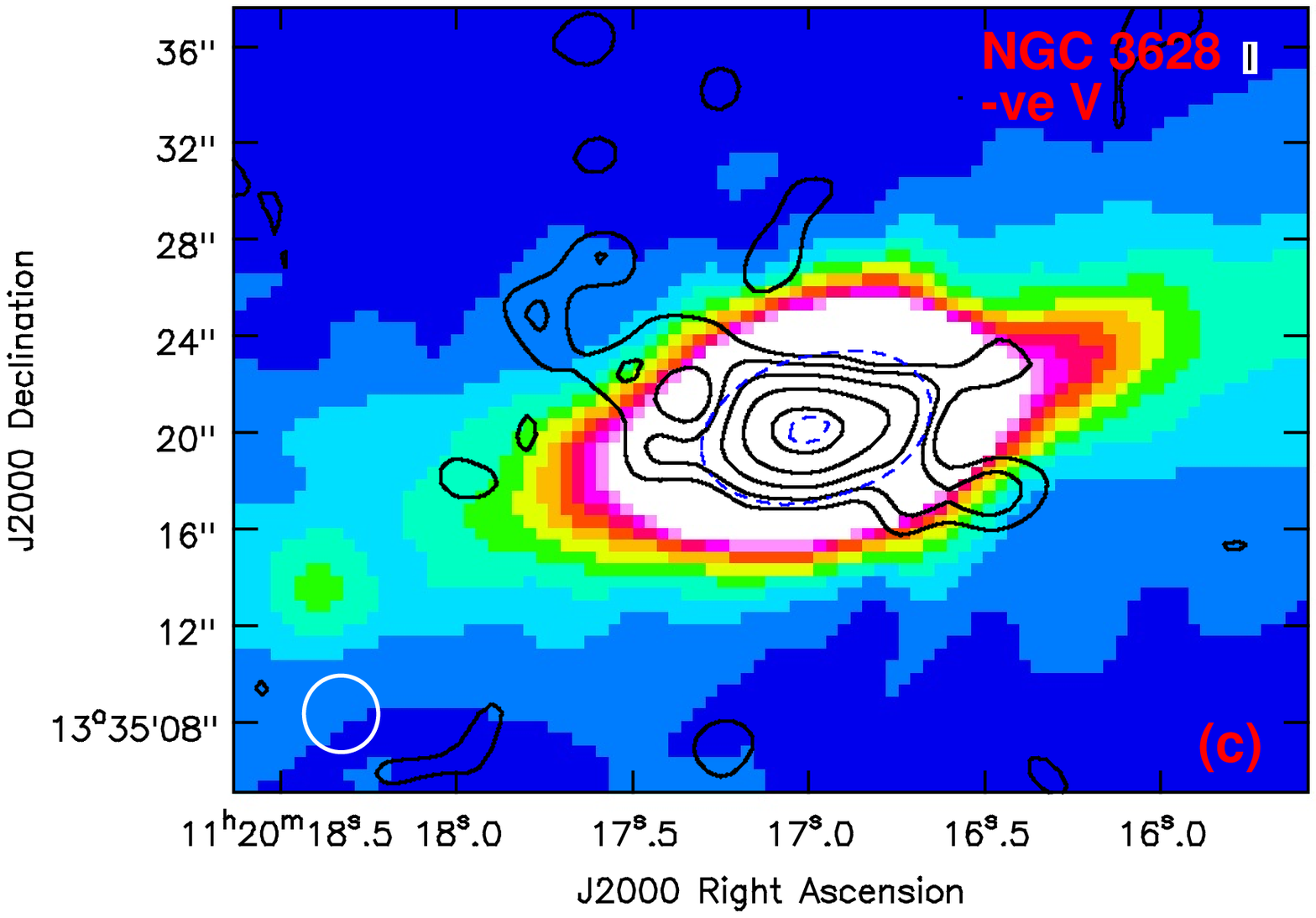}
         \includegraphics*[width=0.45\linewidth]{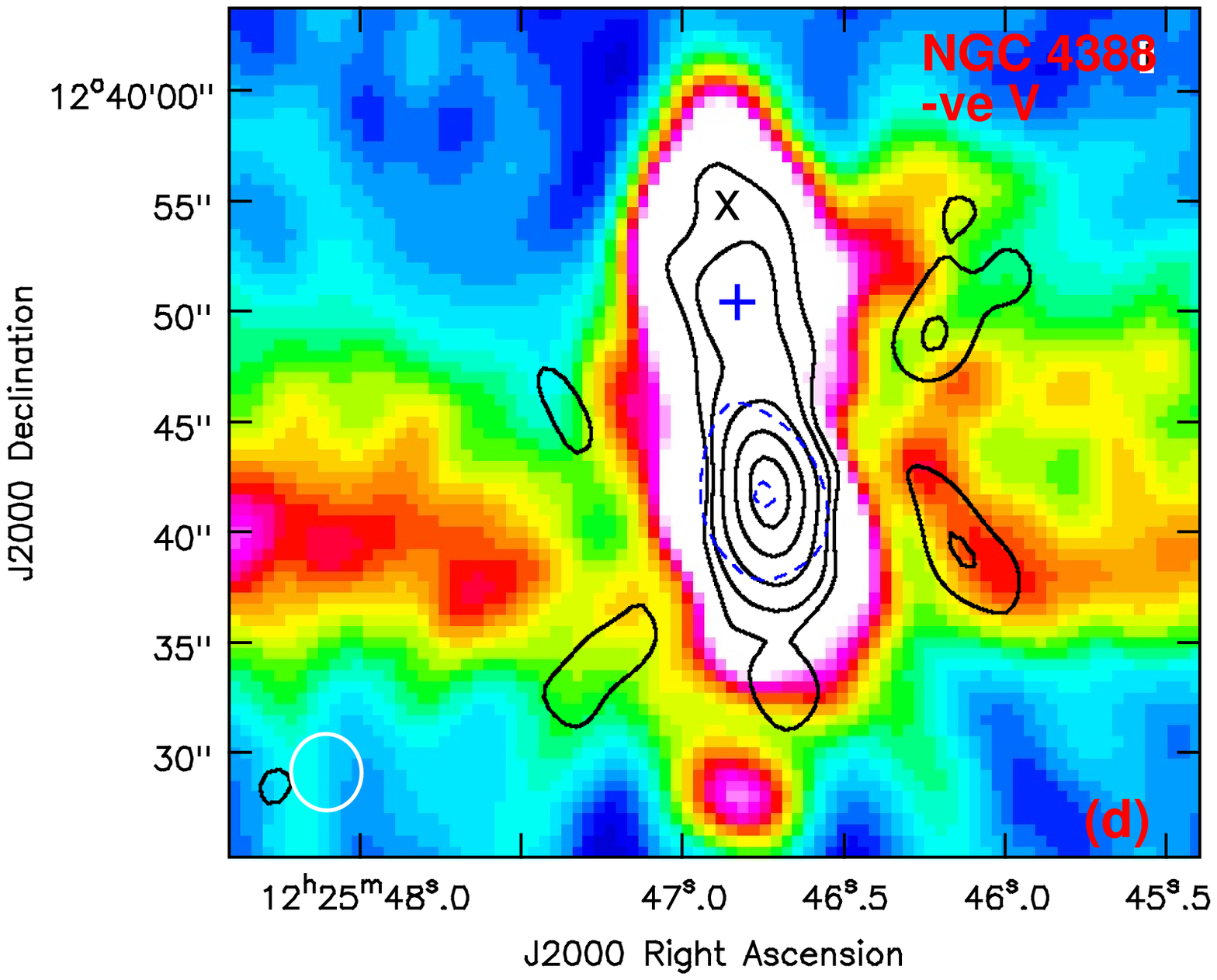}
    \end{subfigure} %
    \begin{subfigure}{}
        \centering
        \includegraphics*[width=0.45\linewidth]{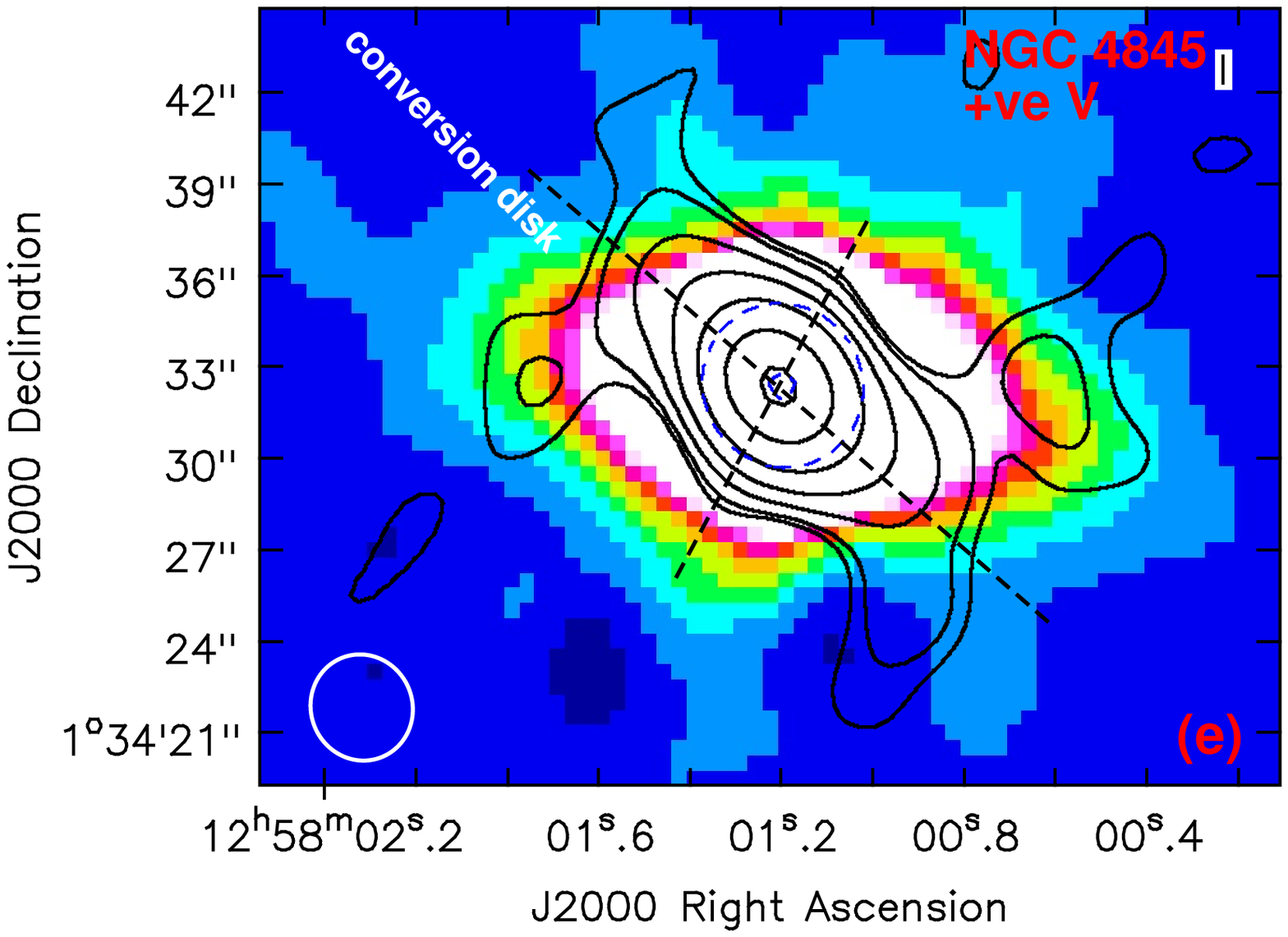}
    \end{subfigure}
    \caption{Galaxies with CP at L-band.  Black contours show the $V$ signal and colour represents the total intensity (labelled I in a small rectangle at upper right), with two dashed contours in blue showing the 95\% and 20\% $I$ levels.  The center of the galaxy is at the center of the 95\% blue dashed contour.  The galaxy name is at upper right with labelling as to whether the CP is positive (+ve) or negative (-ve).  The beam is shown as a white circle at lower left. The first $V$ contour is at 2$\sigma$, where $\sigma$ is measured near the source. Weak residual sidelobes are still evident in a few cases. $V$ contours are: 100, 150, 250, 400, and 600 $\mu$Jy beam$^{-1}$ for NGC~660,
      -38, -75, -130, and -200 $\mu$Jy beam$^{-1}$ for NGC~3079, -27, -45, -75, and -120  $\mu$Jy beam$^{-1}$ for NGC~3628,
      -50, -100, -200, -350, and -500 $\mu$Jy beam$^{-1}$ for NGC~4388, and
      48, 100, 250, 500, 1000, 2000, and 3300 $\mu$Jy beam$^{-1}$ for NGC~4845. Other dashed lines and  annotations are discussed in Sect.~\ref{sec:discussion}.
    } \label{fig:allgalaxies}
\end{figure*}

\subsection{Summary and Intercomparison}
\label{sec:summary_intercomparison}

Let us adopt $\approx$ 0.2\% (Table~\ref{table:detections} and Sect.~\ref{sec:introduction}) as the value at which one {\it could} detect CP, had the calibrator errors and rms not been so high.  If that were the case, then only 9 CHANG-ES galaxies had observations that were sensitive enough to have detected CP (Tables~\ref{table:nondetections} and \ref{table:detections}), of which 5 were actually detected.  
Although the statistics are low, this is a 55\% detection rate, irrespective of any knowledge of the presence or absence of an AGN.  It is tempting to extrapolate and suggest that, with sufficiently sensitive and high fidelity observations, the majority of the galaxies which have AGN would also display CP.  The final count of AGNs in the CHANG-ES sample is left to another paper, but it may very well be that all AGNs in nearby galaxies harbour a circularly polarized component at some time.

A Stokes $V$ signal is, itself, compelling evidence for the presence of an AGN since star forming regions are unable to explain such a CP signal.
There are other galaxies in the CHANG-ES sample that are also known to have AGNs yet we do not detect CP in them.  It isn't always clear why this is the case, but a strong contributor is surely the limitation of high upper limits in the observations (Table~\ref{table:nondetections}).  Another possibility is that CP is known to be highly variable, so applying an across-the-board 0.2\% limit on CP may not reflect the reality of the physical conditions in these sources.

All of our CP galaxies show a hard X-ray source in their cores and all also display further evidence for an AGN (LINER, VLBI detections, etc., Sect.~\ref{sec:discussion}).  {In Table~\ref{table:xrays} we list values for the cores of the CP-detected sources from XMM-Newton data in a consistent fashion and for the same energy range. In Fig.~\ref{fig:xraygraph} we plot the CP signal against the X-ray signal from the galaxy cores, both in Solar luminosity units.  In all cases, the XMM data were taken significantly before the radio data, the shortest time separation being approximately one year.

  We only know of two outbursts: a radio outburst in NGC~660 that occurred between 2008 and 2010.7 and the X-ray outburst of NGC~4845.
  For the remaining sources, we do not have any information about outbursts, but suspect that future monitoring programs may shed more light on whether variability could be a clue that CP will be detected.

  One important conclusion from Fig.~\ref{fig:xraygraph}, however, is that the two galaxies that (by far) have the strongest X-ray cores are also the galaxies that show the strongest $m_C$; both NGC~4388 and NGC~4845 have values of order 2\% in absolute value (Table~\ref{table:detections}). The internal absorption specified in this table is also interesting.  Only NGC~3079 does not show an internal absorption component and this is also our most marginal signal.  On the other hand, NGC~4388 and NGC~4845, our strongest sources, {\it also show the strongest internal absorption}.  {Internal absorption is relatively easy to determine in X-rays and, again, this result suggests a link between the X-ray emission and CP.}

}

In 4 out of 5 cases, CP shows {\it structure}, sometimes clearly jet-like morphology (Fig.~\ref{fig:allgalaxies}). In the fifth case, NGC~3079, is a marginal detection with no structure observed.  This is the first time that CP structures have been observed in nearby galaxies and opens up the possibility of probing not only the core, but jet or accretion disk structures.


\section{Conclusions}
\label{sec:conclusions}

CP has the potential to probe deep into the cores of galaxies and provide information on the magnetic field strength and cosmic ray electron energy distribution.  CP is also a clear indication of the presence of an AGN (or LLAGN) and may be helpful in sorting out whether larger-scale activity is AGN or starburst-related. {
  Since {\it linear} polarization may not be detectable due to Faraday depolarization, CP may in fact be the only way to probe the physical conditions deep into cores at these low frequencies.  Ideally, higher frequency measurements of linear polarization might be extrapolated to the lower frequencies (as we suggest in Sect.~\ref{sec:N4388_conversion})
as long as the connection is clear. If there are multiple components, however, such a connection may not be straightforward.  }

We have carefully searched through the CHANG-ES high resolution (B-configuration) L-band data for CP signals.  Many of the galaxies do not have sufficiently high signal-to-noise and sufficiently low calibration errors to detect CP.   However, we find 5 nearby galaxies with detectable (and sometimes very strong) CP, namely NGC~660, NGC~3079, NGC~3628, NGC~4388, and NGC~4845.  (NGC~3079, however, is only a marginal detection).  To our knowledge, this more than doubles the number of known AGNs with CP-producing central regions in spiral galaxies, the previous ones being only the Milky Way and M~81.

CP is observationally challenging to measure and also is difficult to interpret; however, the most successful mechanism to date is {\it Faraday conversion}.
We show that {\it all} of our detections have steep L-band to C-band spectral indices [$\alpha_V(L\,-\,C) \ltabouteq\,-3$] that are typically expected from Faraday conversion.

We have provided some examples as to how the {\it Faraday conversion} interpretation is most closely aligned with the data  and can lead to estimates of the physical parameters of AGNs (Sect.~\ref{sec:reconciliation}).  Moreover, we have expanded the analysis of \citet{irw15} to provide {\it analytical formulae} for a general $p$, which is the energy spectral index of cosmic ray electrons and is a measurable quantity because it can be determined from total intensity measurements in the optically thin limit (here taken as being C-band). These equations are given in
Eqns.~\ref{convp<2practical}, \ref{convp=2practical}  and \ref{convp>2practical}; Eqn.~\ref{convp=2practical} also supercedes and is more accurate than the earlier equation given in \citet{irw15}.

An unexpected result is that the CP is {\it resolved} for 4 of our sources (Fig.~\ref{fig:allgalaxies}). For the first time, CP is seen in {\it structures} in nearby galaxies.  For example, CP is seen along the northern jet in NGC~4388 at a very high level ($\approx\,3$ \%).  Such measurements present us with a new probe of the physical conditions in nearby jets.

Also, for the first time, we have seen what we have named a {\it conversion disk} in NGC~4845 (Fig.~\ref{fig:allgalaxies}e).
This disk is {\it at an angle} to the galaxy's major axis.  It is roughly perpendicular to an apparent outflow direction \citep[cf.][]{per17} suggesting that the inner part of this disk may be related to an accretion disk that aligns an outflow.  However, the conversion disk is seen {\it only} in CP.  This suggests that an originally linearly polarized signal associated with an extended accretion disk has been converted to CP in a region that has properties adequate for this conversion to occur.  It also suggests that many more LLAGNs in the nearby universe may reveal their accretion disks via CP measurements only.

Given the frequency of detections in the CHANG-ES survey which was not set up for CP measurements, these results raise the possibility that CP may actually be quite a normal, although possibly variable characteristic of AGNs that are embedded in spiral galaxies. 
{Our X-ray measurements also indicate that galaxies with the strongest X-ray core emission are also the galaxies that show the strongest CP ($m_C\,\approx\,2\%$ in absolute value) as well as the strongest internal absorption.  Thus, targetting strong X-ray cores in nearby spiral galaxies could result in a high frequency of CP detections.
  }

\acknowledgments

This work has been supported by a Discovery Grant to the first author by the Natural Sciences and Engineering Research
Council of Canada. YS acknowledges generous support by the Hans-Boeckler Foundation.  The CHANG-ES project at Ruhr-University Bochum has been supported by DFG through FOR1254.
This research has made use of the NASA/IPAC Extragalactic Database (NED) which is operated by the Jet Propulsion Laboratory, California Institute of Technology, under contract with the National Aeronautics and Space Administration. 
The National Radio Astronomy Observatory is a facility of the National Science Foundation operated 
under cooperative agreement by Associated Universities, Inc. 

\newpage

\appendix

\section{Circular polarization by relativistic conversion}\label{appendixA}

In \citet{irw15} (hereafter CHANG-ES V), a useful approximation for the production of circular polarization in a GHz peaked radio source was given. This assumes the dominance of relativistic particles over thermal particles and correspondingly weak Faraday rotation in a magnetized plasma. The mechanism is the differential rotation in the plasma of the two modes in elliptically polarized synchrotron radiation. 
The formulae were given for the special case when the power law energy distribution of the electrons was $E^{-2}$.
See CHANG-ES V Appendix E for details of that development in which we were primarily concerned with the frequency dependence of CP.

We now generalize this result to the energy distribution $E^{-p}$ (i.e. $N(E)\,\propto\,E^{-p}$) of the relativistic electrons. We consider the Faraday thin case since, as we showed in CHANG-ES V, once a source becomes Faraday thick, the conversion term becomes very small (see CHANG-ES V Appendix E for a more technical definition of `Faraday thick' and `Faraday thin' in this context). Thus, along a line of sight, the conversion is occurring mainly prior to Faraday thickness.

The approximate result for the flux density, $S_V$, includes an emission term (subscript $em$) and a conversion term (subscript $conv$) which is given in Eqn. E11 of CHANG-ES V and maintains the form
\be
S_V=  S_{V_{em}}\,+\,S_{V_{conv}}\,=\,\Omega_s \eta_V L + U_o \kappa_c L =    \Omega_s\frac{\tilde\eta_V}{\nu_9}L+\frac{U_o\tilde\kappa_cL}{\nu_9^3},\label{eq:V}
\ee
In this expression $\Omega_s$ is the source solid angle, $U_o$ is the linearly polarized flux entering the plasma medium, $L$ is the line of sight distance through the source, $\nu_9$ is the observed frequency in GHz, $\eta_V$ is the emission coefficient for $S_V$ in the relativistic medium, and $\kappa_c$ is the conversion coefficient from $U_o$ to $S_{V_{conv}}$.
With this formalism, based on \citet{bec02}, the conversion coefficient in Stokes $Q$ is zero so we need only $U_o$ to describe the initial linearly polarized flux.
$\tilde  X$ indicates the value of a quantity, $X$, at one GHz (e.g. $\tilde\eta_V$ is the value of $\eta_V$ at 1 GHz). The coefficients, however, are function of {\it both} frequency as well as $p$.  We now consider these coefficients, $\tilde\kappa_c$ and $\tilde\eta_V$.

The {\it conversion coefficient} \citep{Saz69, bec02} for ultra relativistic particles is given by ($p\ne 2$) 
\be
\tilde\kappa_c=-\pi\frac{\nu_p^2\nu_B^2}{\lambda_9} \sin^2{\theta} f(p,\nu),\label{eq:kappa_c}
\ee
where $\theta$ is the angle between the line of sight and the magnetic field, $B$ (the perpendicular field is $Bsin(\theta)$), and $\lambda_9=c/(10^9 Hz)$. {Note that $\theta$ ranges from $0$ to $<90$ degrees since we only see emission from particles that have a component of velocity moving towards us.}
The squared plasma frequency
\be
\nu_p^2=\frac{n_e e^2}{\pi m_e},\label{eq:plasma_f}
\ee
and the  electron gyro frequency
\be
\nu_B=\frac{eB}{2\pi m_e c}\label{eq:gyro_f}
\ee
are to be measured in GHz. The constants, $e$, $m_e$, and $c$ are the electric charge, the electric mass, and the speed of light, respectively. {Following \citet{bec02}, the electron density  in the plasma frequency, $n_e$, comprises both thermal and non-thermal electrons with the non-thermal component dominating.}  The power law function of Eqn.~\ref{eq:kappa_c}, $f(p,\nu)$, is  ($p\ne 2$) 
\be
f(p,\nu)\equiv \frac{2}{p-2}\left(\frac{\nu_9}{\nu_B\sin\theta}\right)^{(1-p/2)}\left(\left(\frac{\gamma_o^2\nu_B\sin\theta}{\nu_9}\right)^{(1-p/2)}-1\right).\label{eq:pwrfct}
\ee
where $\gamma_o$ is the lower energy cut-off of the relativistic particles. Here and throughout this development, we have set the pitch angle equal to the angle between the line of sight and the magnetic field since, for ultra relativistic particles, they are about the same.

The {\it emission coefficient} may  be written for a general energy power law as \citep[e.g.][their Eqn.~D11]{bec02} 
\be
\tilde\eta_V=-\eta_o  \cos{\theta}(\nu_B\sin{\theta})^{p/2}\nu_9^{(1-p/2)}g(p),\label{eq:eta_V}
\ee
where 
\be
\eta_o\equiv \pi \nu_p^2\nu_B\frac{m_e c^2}{\lambda_9^3}.
\ee
Again, we will measure all frequencies in GHz.  The function $g(p)$ is 
\be
g(p)\equiv \frac{3^{(p-1)/2}(p+2)}{2p}\Gamma\left(\frac{p}{4}+\frac{2}{3}\right)\Gamma\left(\frac{p}{4}+\frac{1}{3}\right)\label{eq:gp}
\ee
where $\Gamma$ represents the gamma function.
These results  assume  the energy distribution function to be in the form (Becker \& Falcke 2002)  
\be
\frac{dN(\gamma)}{d\gamma}=n_{er}\frac{|p-1|}{\gamma_o^{(1-p)}}\gamma^{-p}
\ee
{where $n_{er}$ is the relativistic electron density.}

\vskip 0.2truein

For the conversion term, it is possible to simplify the function $f$ of Eqn.~\ref{eq:pwrfct} under certain conditions. If the peak frequency $\gamma^2 \nu_B$ is used in this equation for $\nu$ and if $\gamma_o \ll \gamma$, then for $p<2$ the first term in the bracket may be ignored.  In that case we see from Eqns.~\ref{eq:kappa_c}  and \ref{eq:pwrfct} that 
\be
\tilde \kappa_c =-\frac{2\pi \nu_p^2\nu_B^2}{(2-p)\lambda_9} \sin^2{\theta}\left(\frac{\nu_9}{\nu_B\sin\theta}\right)^{(1-p/2)},\label{eq:limpwrfct}
\ee
and so  provided that $p<2$, $\kappa_c = \tilde\kappa_c/{\nu_9}^3 \propto \nu_9^{-(2+p/2)}$  rather than $\nu_9^{-3}$. Should $p>2$ be the case and the frequency ratio remain the same, then the first term in the brackets dominates and $f=const$. Hence the $\nu_9^{-3}$ behaviour is maintained. 

We may note finally that if $\nu_9\approx \gamma_o^2\nu_B$, then over a band width $\Delta\nu/\nu\ll 1$ $f$ may increase strongly away from zero as $\nu$ increases. This would be reflected in the sudden onset of circular polarization across the bandwidth and possible sign changes. (An example is given in Fig.~\ref{fig:positive} and discussed in Sect.~\ref{sec:N3628_conversion}.)

The emission term generalizes slightly given Eqn.~\ref{eq:eta_V} to give the frequency dependence 
\be
\eta_V=\frac{\tilde\eta_V}{\nu_9}\propto \nu_9^{-p/2}, \label{eq:eta_Vexp}
\ee
which holds for all $p$.

In summary the formal expression for $S_V\,=\,S_{V_{em}}+S_{V_{conv}}$ in this model takes the following forms for $S_{V_{em}},~S_{V_{conv}}$:

\begin{eqnarray}
  S_{V_{em}}&=&-\Omega_s \eta_o L\cos{\theta}(\nu_B\sin{\theta})^{p/2}\nu_9^{(-p/2)}g(p)~~~~~~~~~(p<2, ~~ p=2,~~p>2)
  \label{emission}
\end{eqnarray}
so the emissive term maintains the same form independent of $p$. However, for the conversion term we have,
\begin{eqnarray}
 &S_{V_{conv}} &- 2\pi U_oL\frac{\nu_p^2 (\nu_B\sin\theta)^{(1+p/2)}}{(2-p)\lambda_9}\left(\frac{1}{\nu_9}\right)^{(2+p/2)}\left[1-\left(\frac{{\gamma_0}^2\nu_B\sin\theta}{\nu_9}\right)^{1-p/2}  \right]~~~~~~~(p<2)
\label{convp<2}
\end{eqnarray}
with the second term in the square brackets only becoming important if $\gamma_o\nu_B\sin\theta \approx \nu_9$.  As noted above, if  $\gamma_o\nu_B\sin\theta < \nu_9$, then that term is negligible and  $S_{V_{conv}}\,\propto\,\nu_9^{-(2+p/2)}$. The equation is the same if $p>2$ but we express it somewhat differently since it is convenient to have $p/2 -1 >0$,
\begin{eqnarray}
  S_{V_{conv}}\,=\,- 2\pi U_oL\frac{\nu_p^2(\nu_B\sin\theta)^{(1+p/2)}}{(p - 2)\lambda_9}\left(\frac{1}{\nu_9}\right)^{(2+p/2)}
  \left[
    \left(
    \frac{\nu_9}{{\gamma_0}^2\nu_B\sin\theta}
  \right)^{(p/2-1)} - 1
  \right]~~~~~(p>2)
  \label{convp>2}
  \end{eqnarray}
Here the second term in the square brackets (-1) becomes important only if the first term $\approx 1$.  As noted above, if $\gamma_o\nu_B\sin\theta < \nu_9$ then $S_{V_{conv}}\,\propto\,\nu_9^{-3}$. For the limiting case, $p=2$,
\begin{eqnarray}
  S_{V_{conv}}\,=\,2\pi U_o L\frac{ {\nu_p}^2\left(\nu_B\sin\theta\right)^2}{\lambda_9}\,\frac{1}{{\nu_9}^3}\,
  ln\left(\gamma_o\sqrt{\left( \frac{\nu_B\sin\theta}{\nu_9} \right)}   \right)~~~~~~(p=2)
  \label{convp=2}
  \end{eqnarray}
in units consistent with Eqns.~\ref{convp<2} and \ref{convp>2}.
This equation is equivalent to the result given in Appendix E of CHANG-ES V except that, in the earlier development, we were primarily concerned with the frequency dependence and so did not include the weak logarithmic term which we now include, as well as a factor, $\pi$.  $B$ in that equation was taken to be the perpendicular field which is $Bsin(\theta)$ here.

\vskip 0.2truein

We now provide practical numerical expressions:
\vskip 0.2truein

{For our estimates, we take the Stokes parameter, $U_0$, as an input value to the relativistic conversion region.  It may be that there is a purely relativistic core producing this value interior to the conversion region.  However, it would be more compelling to produce it internally.  Such linear polarization can be generated by Faraday rotation or a rotating magnetic field as revealed in \citet{jon88} but does not occur in our homogeneous magnetic field with a small Faraday rotation.

  In the remainder of this Appendix, $n_e$ can effectively be taken as $n_{er}$ since relativistic particles dominate.}

\noindent
  For $S_{V_{em}}$ (mJy), with  $L$ (pc), $\nu_9$ (GHz), $B$ (Gauss), $n_e$ (cm$^{-3}$), $\nu_9$ (GHz), and the source size, $\phi$ (arcsec),
\begin{eqnarray}
  S_{V_{em}}&=& -122\,\phi^2 \,L\,n_e B \cos\theta\left(
  \frac{2.8\times 10^6 B\sin\theta}{\nu_9}
  \right)^{p/2} \,g(p)~~~~~~~~~({\rm any}~p)
  \label{emission}
  \end{eqnarray}
This equation corrects the magnitude of the emission term in Appendix E of CHANG-ES V. If $p=2$, then the frequency dependence goes as $\nu^{-1}$, as found previously.  {Adopting values similar to} those of NGC~4845, if $\phi=0.001$, $L=0.03$, $n_e=100$, $B=0.04$, $\nu_9=1.5$, $\theta=\pi/4$ and $p=2$, we find
$S_{V_{em}}\approx 1$ mJy (in absolute value). 

\vskip 0.2truein
\noindent For $S_{V_{conv}}$ (mJy), with $L$ (pc), $\nu_9$ (GHz), $B$ (Gauss) and $n_e$ (cm$^{-3}$), where $U_o$ (mJy) is the linearly polarized flux density to be converted:
\begin{eqnarray}
  S_{V_{conv}}&=& -\frac{5.21\,\times\,10^{7} U_o \,L \,n_e\, (0.0028 B\sin\theta)^{1+p/2}\,{\nu_9}^{-(2+p/2)}}{2-p}\left[
1\,-\, \left(\frac{{0.0028\gamma_o}^2B\sin\theta}{\nu_9}\right)^{1-p/2}
\right]~(p<2)\nonumber\\\label{convp<2practical}\\
  S_{V_{conv}}&=& 4.078\,\times\,10^{2} U_o \,L \,n_e\, (B\sin\theta)^{2}\,{\nu_9}^{-3}ln\left[
  0.053 \sqrt{
\frac{{\gamma_o}^2B\sin\theta}{\nu_9}
  }
  \right] ~~~~~~~~~~~~~~~~~~~~~~~~~~~~~~~~~~~~~(p=2)\nonumber\\
  \label{convp=2practical}\\
    S_{V_{conv}}&=& -\frac{5.21\,\times\,10^{7} U_o \,L \,n_e\, (0.0028 B\sin\theta)^{1+p/2}\,{\nu_9}^{-(2+p/2)}}{p-2}
  \left[
\left(\frac{\nu_9}{0.0028{\gamma_o}^2 B\sin\theta}\right)^{p/2-1} - 1
    \right]
  ~(p>2)\nonumber\\\label{convp>2practical}
\end{eqnarray}
Note that in the text (Sect.~\ref{sec:pne2} and throughout) we let
\begin{equation}
x\,\equiv\, \frac{0.0028{\gamma_0}^2 B\,\sin\theta}{\nu_9}
  \end{equation}
 The combination of all terms that multiply $U_o$ must be $< 1$ because  $S_{V_{conv}}$ must be $< U_o$ (in absolute value).  The above equations are not valid for a combination of terms $> U_o$ since this development assumes that the Stokes $V$ opacity is small (see CHANG-ES V Appendix E for a more precise definition).

 For future reference, we isolate the frequency-dependent terms in Eqn.~\ref{convp>2practical} and define the function
 \begin{equation}\label{modulating}
   F_{\nu_9}\,=\,{\nu_9}^{-(2+p/2)}
  \left[
\left(\frac{\nu_9}{0.0028{\gamma_o}^2 B\sin\theta}\right)^{p/2-1} - 1
    \right]
 \end{equation}
 This is the function that is plotted in Fig.~\ref{fig:positive}.

 Again, {adopting values similar to those found for NGC~4845}, for $L=0.01$ pc, $n_e=100$ cm$^{-3}$, $B=0.04$ Gauss, $\gamma_o=100$, $\theta=\pi/4$, and $\nu_9=1.5$, then $S_V/U_0 = -0.34$ for $p=1.5$, $S_V/U_0 = -0.031$ for $p=2.0$, and $S_V/U_0 = -0.0029$ for $p=2.5$. Strong and significant changes, including sign changes, can occur as the two terms in square brackets for $p<2$, $p>2$ become comparable.    
 The logarithmic term of Eqn.~\ref{convp=2practical} can also be positive or negative. The result is, of course, entirely dependent on the magnitude of the linearly polarized flux to be converted so, for example, a polarized flux of 50 mJy could produce a circularly polarized flux of 1.6 mJy using the above values if $p=2$.
 The strongest conversion, however, occurs for $p<2$ (corresponding to a flatter electron energy distribution) in the above approximations. A linearly polarized flux of 50 mJy could produce 17 mJy of CP if $p=1.5$ in this example.  As has been found by others, the conversion term likely outweighs the emission term in most real situations.

 {Finally, a comment on the sign of the CP. It is important to note that Faraday conversion depends on the perpendicular component of the magnetic field, as opposed to the line-of-sight component that is important for normal Faraday rotation.  Conversion does not depend on the sign of the charge nor on the orientation of the B-field as long as it has a transverse component, as noted in \citet{bec03}. It is the sign of Stokes $U$ to begin with and the conversion depth, $\kappa_c L$ (Eqn.~\ref{eq:V}) that determines the sign.
   }

\vskip 2truein

{\it Facilities:} \facility{VLA}.


\begin{thebibliography}{}
\bibitem[Aller, Aller, \& Plotkin (2003)]{all03} Aller, H. D., Aller, M. F., \& Plotkin, R. M. 2003, \apss, 288, 17
\bibitem[Alton et al. (1998)]{alt98}Alton, P.B., Trewhella M., Davies J.I., Evans R., et al. 1998, \aap, 335, 807
\bibitem[Argo et al. (2015)]{arg15} Argo, M. K., van Bemmel, I. M., Connolly, S. D., \& Beswick, R. J. \mnras, 452, 1081 
\bibitem[Arnaud (1996)]{arn96} Arnaud, K. A. 1996, in Jacoby, G., Barnes, J., eds,  ASP Conf. Ser. Vol. 101, 
Astronomical Data Analysis Software and Systems. Astron. Soc. Pac., San Francisco, p. 17
\bibitem[Asmus et al. (2014)]{asm14} Asmus, D., Honig, S. F. H., Gandhi, P., Smette, A., \& Duschl, W. J. 2014, \mnras, 439, 1648
\bibitem[Beckert (2003)]{bec03} Beckert, T. 2003, \apss, 288, 123
\bibitem[Beckert \& Falcke (2002)]{bec02} Beckert, T. \& Falcke, H. 2002, \aap, 388, 1106.
\bibitem[Bhatnagar et al. (2008)]{bha08} Bhatnagar, S., Cornwell, T. J., Golap, K., \& Uson, J. M. 2008, \aap, 487, 419
\bibitem[Braine et al. (1993)]{bra93} Braine, J., Combes, F., Casoli, F. 1993, \aaps, 97,887
\bibitem[Bower et al. (2002)]{bow02}Bower, G. C., Falcke, H., Sault, R. J., \& Backer, D. C. 2002, 571, 843
\bibitem[Bower (2003)]{bow03}Bower, G. C. 2003, \apss, 288, 69
\bibitem[Briggs (1995)]{bri95} Briggs, D. S. 1995, {\it High Fidelity Deconvolution of
Moderately Resolved Sources}, PhD Thesis, The New Mexico Institute of Mining and Technology, Socorro, NM
\bibitem[Brisken (2003)]{bri03} Brisken, W. 2003, EVLA Memo 58, {\tt https://library.nrao.edu/public/memos/evla/legacy/evlamemo58.pdf}
\bibitem[Brunthaler et al. (2001)]{bru01}Brunthaler, A., Bower, G. C., Falcke, H., \& Mellon, R. R. 2001, \apj, 560, L123
\bibitem[Brunthaler, Bower, \& Falcke (2006)]{bru06}Brunthaler, A., Bower, G. C., \& Falcke, H. 2006, \aap, 451, 845
\bibitem[Carter \& Read (2007)]{car07} Carter, J. A., \& Read, A. M. 2007, \aap, 464, 1155
\bibitem[Cecil, Bland-Hawthorn, \& Veilleux (2002)]{cec02} Cecil G., Bland-Hawthorn J., \& Veilleux S., 2002, \apj, 576, 745
  \bibitem[Condon (1992)]{con92}Condon, J. J. 1992, \araa, 30, 575
\bibitem[Cornwell, Golap \& Bhatnagar (2008)]{cor08b} Cornwell, T. J., Golap, K., \& Bhatnagar, S. 2008,
  IEEE J. of Selected Topics in Signal Proc., Vol. 2, No. 5, 647
\bibitem[Damas-Segovia et al. (2016)]{dam16} Damas-Segovia, A., Beck, R., Vollmer, B., et al. 2016, \apj, 824, 30
  (CHANG-ES VII)
\bibitem[de Bruyn (1977)]{deb77} de Bruyn, A. G. 1977, \aap, 58, 221
\bibitem[Dong \& De Robertis (2006)]{don06} Dong, X. Y., \& De Robertis, M. M. 2006, \aj, 131, 1236
\bibitem[En{\ss}lin (2003)]{ens03} En{\ss}lin, T. 2003, \apss, 288, 183
\bibitem[Everett \& Weisberg (2001)]{eve01} Everett, J. E., \& Weisberg, J., M. 2001, \apj, 553, 341
\bibitem[Filho, Barthel, \& Ho (2002)]{fil02} Filho, M. E., Barthel, P. D., \& Ho, L. C. 2002, \aaps, 142, 223  
\bibitem[Filho et al. (2004)]{fil04} Filho, M., Fraternali, F., Markoff, S., et al. 2004, \aap, 418, 429
\bibitem[Flohic et al. (2006)] {flo06} Flohic, H. M. L. G., et al. 2006 \apj, 647, 140
\bibitem[Gabriel et al. (2004)]{gab04} Gabriel, C., Denby, M., Fyfe, D. J., et al. 2004, ASPC, 314, 759
\bibitem[Gabuzda (2003)]{gab03}Gabuzda, D. C. 2003, \apss, 288, 39
\bibitem[Giroletti \& Panessa (2009)]{gir09} Giroletti, M., \& Panessa, F. 2009, \apj, 706, L260 
\bibitem[Gonz{\'a}lez-Mart{\'i}n et al. (2006)]{gon06} Gonz{\'a}lez-Mart{\'i}n, O. 2006, \aap, 460,45
\bibitem[Goulding \& Alexander (2009)]{gou09} Goulding, A. D., \& Alexander, D. M. 2009, MNRAS, 398, 1165
\bibitem[Herrnstein et al. (2005)]{her05} Herrnstein, J. R., Moran, J. M., Greenhill, L. J., \& Trotter, A.S.
  2005, \apj, 629, 719
\bibitem[Homan \& Wardle (1999)]{hom99} Homan, D. C., \& Wardle, J. F.C. 1999, \aj, 118, 1942
\bibitem[Homan et al. (2001)]{hom01} Homan, D. C, Attridge, J. M., \& Wardle, J. F. C. 2001, \apj, 556, 113
\bibitem[Homan \& Lister (2006)]{hom06} Homan, D. C., \& Lister, M. L. 2006, \aj, 131, 1262
\bibitem[Homan \& Wardle (2003)]{hom03} Homan, D. C., \& Wardle, J. F. C. 2003, \apss, 288, 29
\bibitem[Homan \& Wardle (2004)]{hom04} Homan, D. C., \& Wardle, J. F. C. 2004, \apj, 602, L13
\bibitem[Irwin et al. (2012)]{irw12a} Irwin, J. A., Beck, R., Benjamin, R. A., et al. 2012a, AJ, 144, 43
(CHANG-ES I)
\bibitem[Irwin et al. (2012b)]{irw12b} Irwin, J. A., Beck, R., Benjamin, R. A., et al. 2012b, AJ, 144, 44
(CHANG-ES II)
\bibitem[Irwin et al. (2013)]{irw13} Irwin, J. A., Krause, M., English, J., et al. 2013, \aj, 146, 164
  (CHANG-ES III)
\bibitem[Irwin et al. (2015)]{irw15} Irwin, J. A., Henriksen, R., Krause, M., et al. 2015, \apj, 809, 172 (CHANG-ES V)
\bibitem[Irwin et al. (2017)]{irw17} Irwin, J. A., Schmidt, P., Damas-Segovia, A., et al. 2017, \mnras, 464, 1333 (CHANG-ES VIII)
\bibitem[Iwasawa et al. (2003)]{iwa03} Iwasawa, K., wilson, A. S., Fabian, A. C., \& Young, A. J. 2003, \mnras, 345, 369
\bibitem[Iyomoto et al. (2001)]{iyo01} Iyomoto N., Fukazawa Y., Nakai N., \& Ishihara Y., 2001, \apj, 561, L69
\bibitem[Jones \& O'Dell (1977)]{jon77} Jones, T. W., \& O'Dell, S. L. 1977, \apj, 214, 522
\bibitem[Jones \& O'Dell (1977a)]{jon77a} Jones, T. W., \& O'Dell, S. L. 1977, \apj, 215, 236
\bibitem[Jones (1988)]{jon88} Jones, T.W. 1988, \apj, 332, 678
\bibitem[Kaastra (1992)]{kaa92} Kaastra, J. S. 1992, {\it An X-Ray Spectral Code for Optically Thin Plasmas} (Internal SRON-Leiden Report, updated version 2.0)
\bibitem[Kalberla et al. (2005)]{kal05} Kalberla, P. M. W., Burton, W. B., Hartmann, D., et al. 2005, \aap, 440, 775
\bibitem[Kondratko, Greenhill, \& Moran (2005)]{kon05} Kondratko, P. T., Greenhill, L. J., \& Moran, J. M. 2005, \apj, 618, 618
\bibitem[Krivonos et al. (2015)]{kri15} Krivonos, R., Tsygankov, S., Lutovinov, A., et al. 2015, \mnras, 448, 3766
\bibitem[LaMassa et al. (2015)]{lam15} LaMassa, S., et al. 2015, arXiv:1412.2136 (accepted for publication in \apj)
\bibitem[Levenberg (1944)]{lev44} Levenberg, K. 1944, Quarterly of Applied Mathematics, 2, 164
\bibitem[Li et al. (2011)]{li11} Li, F., Brown, S., Cornwell, T. J., \& de Hoog, F. 2011, \aap, 531, A126
\bibitem[Li et al. (2016)]{li16} Li, J.-T., Beck, R., Dettmar, R.-J., et al. 2016, \mnras, 456, 1723 (CHANG-ES VI)
\bibitem[Macquart (2002)]{mac02} Macquart, J-P. 2002, \pasa, 19, 43
\bibitem[Marcquart (1963)]{mar63} Marcquart, D. W. 1963, Journal of the Society for Industrial and Applied Mathematics, 11(2), 431
\bibitem[McKinney, Tchekhovskoy, \& Blandford (2013)]{mck13} McKinney, J. C., Tchekhovskoy, A., \& Blandford, R. D. 2013, Science, 339, 49  
\bibitem[McMullin et al. (2007)]{mcm07} McMullin, J. P., et al. 2007, Astronomical Data Analysis Software and Systems
  XVI, ASP Conf. Series, Vol. 376, Ed., R. A. Shaw, F. Hill, \& D. J. Bell, p. 127
\bibitem[Mewe et al. (1985)]{mew85} Mewe, R., Gronenschild, E. H. B. M., \& van den Oord, G. H. J. 1985, \aaps, 62, 197
\bibitem[Minchin et al. (2013)]{min13} Minchin R. F., Ghosh T., Momjian E., Salter C. J., 2013, in
American Astronomical Society Meeting Abstracts, Vol.
221, American Astronomical Society Meeting Abstracts,
p. 157.06
\bibitem[Middelberg et al. (2007)]{mid07} Middelberg, E., Agudo, I., Roy, A. L., \& Krichbaum, T. P. 2007, \mnras, 377, 731
\bibitem[M{\"u}ller, Beck \& Krause (2017)]{mul17} M{\"u}ller, P., Beck, R., \& Krause, M. 2017, \aap, 600, 63
\bibitem[Mu{$\tilde {\rm n}$}oz et al. (2012)]{mun12}
	Mu{$\tilde {\rm n}$}oz, D. J., Marrone, D. P., Moran, J. M., \& Rao, R. 2012, \apj, 745, 115
\bibitem[Nikolajuk \& Walter (2013)]{nik13} Nikolajuk, M. \& Walter, R. 2013, \aap, 552,A75 
\bibitem[Myserlis (2015)]{mys15}Myserlis, I. E. 2015, PhD thesis, Max-Planck-Institut  f{\"u}r Radioastronomie,
({\tt http://kups.ubi.uni-koeln.de/6967/})
\bibitem[Myserlis et al. (2017)]{mys17} Myserlis, I. E. Angelakis, E.,  Kraus, A., et al. 2017, arXiv:1706.04200
\bibitem[O'Sullivan et al. (2013)]{osu13} O'Sullivan, S.P., McClure-Griffiths, N.M., Feain, I.J., et al. 2013,\mnras,435,311
\bibitem[Pacholczyk (1977)]{pac77} Pacholczyk, a. G. 1977, {\it Radio Galaxies}, Pergamon Press, Oxford
\bibitem[Pearson \& Readhead (1984)]{pea84}Pearson, T. J., \& Readhead, A. C. S. 1984, \araa, 22.97
\bibitem[Perley (2016)]{per16} Perley, R. A. 2016, EVLA Memo 195, {\tt https://library.nrao.edu/public/memos/evla/EVLAM\_195.pdf}
\bibitem[Perley \& Butler (2013)]{per13} Perley, R. A., \& Butler, B. J. 2013, \apjs, 204, 19
\bibitem[Perley \& Butler (2017)]{perley17} Perley, R. A., \& Butler, B. J. 2017, \apjs, 230, 7
\bibitem[Perlman et al. (2017)]{per17} Perlman, E. S., Meyer, E. T., Wang, Q. D., et al. 2017, \apj, 842, 126
\bibitem[Ptak (2001)]{pta01} Ptak, A. 2001, in {\it X-RAY ASTRONOMY: Stellar Endpoints,AGN, and the Diffuse X-ray Background}. Ed. N. E. White, G. Malaguti, and G. G.C. Palumbo, Melville, NY: American Institute of Physics, 2001. AIP Conference Proceedings, 599, p. 326
\bibitem[Rau \& Cornwell (2011)]{rau11} Rau, U., \& Cornwell, T. J. 2011, \aap, 532, A71
\bibitem[Rayner, Norris, \& Sault (2000)]{ray00} Rayner, D. P., Norris, R. P., \& Sault, R. J. 2000, \mnras, 319, 484
\bibitem[Sazonov (1969)]{Saz69} Sazonov, V.N. 1969, Soviet Astr., 13, 396
\bibitem[Sault \& Perley (2013)]{sau13} Sault, R. J., \& Perley, R. A. 2013, EVLA Memo 170, {\tt https://library.nrao.edu/public/memos/evla/EVLAM\_170.pdf}
\bibitem[Shafi et al. (2015)]{sha15} Shafi, N., Oosterloo, T. A., Morganti, R., et al.
  2015, \mnras, 454, 1404
\bibitem[She et al. (2017)]{she17}She, R., Ho, L. C., \& Feng, H. 2017, \apj, 835, 223
\bibitem[Simmons \& Stewart (1985)]{sim85}Simmons, J. F. L., \& Stewart, B. G., \aap, 142, 100
\bibitem[Stevens, Amure, \& Gear (2005)]{ste05}Stevens, J. A., Amure, M., \& Gear, W. K. 2005, \mnras, 357, 361
\bibitem[Str{\"u}der et al. (2001)]{str01} Str{\"u}der, L., Briel, U., Dennerl., K., et al. 2001, \aap, 365, 18
\bibitem[Tsai et al. (2012)]{tsa12} Tsai, A.-L., et al. 2012, \apj, 752, 38
\bibitem[Turner et al. (2001)]{tur01} Turner, M. J. L., Abbey, A., Arnaud, M., et al. 2001, \aap, 365, 27
\bibitem[Vaillancourt (2006)]{vai06} Vaillancourt, J. E. 2006, \pasp, 118, 1340
\bibitem[van Driel et al. (1995)]{vand95} van Driel, W., Combes, F., Casoli, F., et al. 1995, \aj, 109, 942
\bibitem[V{\'e}ron-Cetty \& Veron (2006)]{ver06} V{\'e}ron-Cetty, M.-P., \& V{\'e}ron, P. 2006, \aap, 455, 773
\bibitem[Wiegert et al. (2015)]{wie15} Wiegert, T., Irwin, J., Miskolczi, A., et al. 2015, \aj, 150, 81 (CHANGES-IV)
\bibitem[Wong et al. (2016)]{won16} Wong, O. I., Koss, M. J., Schawinski, K., et al. 2016, \mnras, 460, 1588
\bibitem[Yaqoob et al. (1995)]{yaq95} Yaqoob, T., et al. 1995, \apj, 455, 508 
\end{thebibliography}
\end{document}